\documentclass[useAMS,usenatbib]{mn2e}
\usepackage{aas_macros,graphicx,times,multirow,amsmath,bbold}
\usepackage[]{color}
\usepackage[draft]{hyperref}

\title[X-ray spectra and polarization from magnetars]{X-ray spectra and polarization from magnetar candidates}
\author[R. Taverna, R. Turolla, V. Suleimanov, A. Y. Potekhin, S. Zane]{R. Taverna$^{1,2}$\thanks{E-mail:
\href{mailto:taverna@fis.uniroma3.it}{taverna@fis.uniroma3.it}}, R. Turolla$^{2,3}$, V. Suleimanov$^{4,5,6}$,
A. Y. Potekhin$^{7}$, S. Zane$^{3}$\\
$^1$Department of Mathematics and Physics, University of Roma Tre, via della Vasca Navale 84, I-00146 Roma, Italy\\
$^2$Department of Physics and Astronomy, University of Padova, via Marzolo 8, I-35131 Padova, Italy\\
$^3$Mullard Space Science Laboratory, University College London, Holmbury St. Mary, Surrey, RH5 6NT, UK\\
$^4$Institut f\"{u}r Astronomie und Astrophysik, Sand 1, D-72076 T\"{u}bingen, Germany\\
$^5$Kazan (Volga region) Federal University, Kremlevskaja str., 18, Kazan 420008, Russia\\
$^6$Space Research Institute of the Russian Academy of Sciences, Profsoyuznaya Str. 84/32, Moscow 117997, Russia\\
$^7$Ioffe Institute, Politekhnicheskaya 26, 194021, Saint Petersburg, Russia
}

\date{Accepted \ldots. Received \ldots; in
original form \ldots} \pagerange{\pageref{firstpage}--\pageref{lastpage}} \pubyear{2019}

\def\LaTeX{L\kern-.36em\raise.3ex\hbox{a}\kern-.15em
    T\kern-.1667em\lower.7ex\hbox{E}\kern-.125emX}

\def\flux {\mbox{erg cm$^{-2}$ s$^{-1}$}}

\def\thetaB {\theta_\mathrm{B}}
\def\thetak {\theta_\mathrm{k}}
\def\Bp {B_{\rm p}}
\def\Deltaphins {\Delta\phi_{\rm N-S}}
\def\phik {\phi_\mathrm{k}}
\def\kvec {\boldsymbol{k}}

\def\nvec {\boldsymbol{n}}

\begin{document}

\label{firstpage}
\maketitle
\begin{abstract}
Magnetars are believed to host the strongest magnetic fields in the 
present universe ($B\ga10^{14}$ G) and the study of their persistent 
emission in the X-ray band offers an unprecendented opportunity to gain insight 
into physical processes in the presence of ultra-strong magnetic 
fields. Up to now, most of our knowledge about magnetar sources came from
spectral analysis, which allowed to test the resonant Compton scattering 
scenario and to probe the structure of the star magnetosphere. On 
the other hand, radiation emitted from magnetar surface is expected 
to be strongly polarized and its observed polarization pattern 
bears the imprint of both scatterings onto magnetospheric charges
and QED effects as it propagates in the magnetized vacuum around
the star. X-ray polarimeters scheduled to fly in the next years
will finally allow to exploit the wealth of information stored
in the polarization observables. Here we revisit the problem of
assessing the spectro-polarimetric properties of magnetar persistent
emission. At variance with previous investigations, proper account
for more physical surface emission models is made by considering
either a condensed surface or a magnetized atmosphere. Results
are used to simulate polarimetric observations with the forthcoming
Imaging X-ray Polarimetry Explorer ({\em IXPE}). We find that X-ray 
polarimetry will allow to detect QED vacuum effects for all the emission 
models we considered and to discriminate among them.
\end{abstract}
\begin{keywords}
polarization -- radiative transfer -- scattering -- stars: magnetars -- techniques: polarimetric -- X-rays: stars
\end{keywords}

\section{Introduction}
\label{intro}
Anomalous X-ray pulsars (AXPs) and soft-gamma repeaters (SGRs) 
stand up as peculiar objects amongst the varied landscape of 
neutron star (NS) sources. The measured spin periods ($P\approx
2$--$12$ s) and spin-down rates ($\dot{P}\approx10^{-15}$--$10^
{-10}$ ss$^{-1}$) set them apart from the bulk of the NS population, 
and point to ultra-strong dipole magnetic fields (up to $10^
{14}$--$10^{15}$ G), orders of magnitude higher than those of
ordinary NSs. Nowadays, a wide consensus gathered in favour 
of the presence of an ultra-magnetized NS, a magnetar, in AXPs/SGRs,
following the original suggestion by \citet{dt92} and \citet[see 
\citealt{tzw15,kb17} for recent reviews]{td95}.
Observationally, AXPs and SGRs manifest themselves through the 
emission of short X-ray bursts (with luminosities $L_{\rm X}\approx 
10^{36}$--$10^{43}$ ergs$^{-1}$ and duration $\Delta t\approx 
0.01$--$50$ s) and, exceptionally, of much more energetic giant 
flares ($L_{\rm X}\approx10^{44}$--$10^{47}$ ergs$^{-1}$ and 
$\Delta t\approx100$--$1000$ s). Presently, the list of magnetar 
sources totals more than 20 objects (plus some candidates; 
\citealt{ok14}\footnote{See the McGill magnetar catalog: \\
http://www.physics.mcgill.ca/\textasciitilde pulsar/magnetar/main.html
\label{webmcgill}}), many of which are transients, detectable 
only during their outburst phases, when the flux increases up 
to a factor $\approx 10^3$ above the quiescent level \cite[see 
e.g.][]{re11}.

Magnetars exhibit also a persistent emission in the $\approx
0.5$--$200$ keV band, with $L_{\rm X}\approx10^{31}$--$10^{36}$ 
ergs$^{-1}$, usually in excess of the rotational energy loss rate, 
$\dot{E}_{\rm rot}\sim4\times10^{46}\dot{P}/P^3$ ergs$^{-1}$. In
the soft X-rays ($0.1$--$10$ keV), the spectrum is typically the 
superposition of a thermal component ($kT\approx1$ keV) and a power-law 
tail (photon index $\Gamma\approx2$--$4$), although transient sources 
often show a purely thermal spectrum. Up to now, many of the magnetar 
emission properties have been unveiled by investigating their spectra; 
this also allowed to shed light on how thermal photons are reprocessed 
in the star magnetosphere, supporting the Resonant Compton Scattering 
(RCS) paradigm (see Thompson, Lyutikov \& Kulkarni \citeyear{tlk02}).


\begin{figure*}
\begin{center}
\includegraphics[width=13cm]{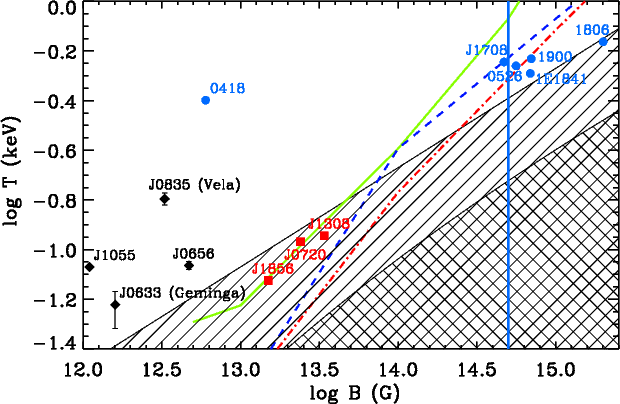}
\caption{The critical temperature for magnetic condensation
as a function of the surface field. The lines refer to
different chemical compositions: iron (green, solid),
carbon (blue, dashed) and helium (red, dash-dotted),
according to the calculations by \citet{ml06,ml07}. 
The shaded and hatched areas mark the condensation 
regions for Fe and H, respectively, according to the 
older estimate by \citet{lai01}. Samples of different 
NS sources are marked with different symbols: cyan filled circles (magnetars;
see \citealt{ok14}\textsuperscript{\ref{webmcgill}}), 
red squares and black diamonds (X-ray dim isolated
NSs and isolated pulsars, respectively; see \citealt{tur09}).
The thick vertical line is drawn in correspondence
to a magnetic field of $5\times10^{14}$ G.}

\label{figure:tcrit}
\end{center}
\end{figure*}
The extremely strong magnetic field near the surface of a magnetar
implies that radiation is highly polarized. This is due to the effects 
of the magnetic field on the optical properties of both the plasma 
and the vacuum in which the photons propagate \cite[see e.g.][for 
a review]{hl06}. Polarimetric observations in the X-rays are bound 
to add a new dimension to our knowledge of magnetars, by providing, 
among others, an independent estimate of the magnetic field strength 
and topology and by unveiling the source geometry \cite[][]{fd11,tav+14,tt17}. 
At variance with spectral measurements, the wealth of information 
encoded in polarization observables is still to be exploited. In 
fact, previous efforts with the {\it OSO-8} satellite \cite[][]{weiss+78} 
and baloon experiments \cite[e.g. {\it PoGO}/{\it PoGOLite};][]{lp04,kp07}
were limited only to a couple of very bright sources with large 
enough polarization degree. Full-fledged polarimetric missions finally 
started to be planned in the last decade and culminated in the 
{\it IXPE} observatory \cite[][]{weiss+16}, a NASA-ASI SMEX mission 
scheduled to fly in 2021, and in the {\em eXTP} satellite \cite[][]{zhang+19},
approved by the Chinese Academy of Science.

Goal of this paper is to provide a systematic assessment of 
the spectral and polarization properties of magnetar persistent
emission in the $2$--$10$ keV band, the working energy range 
of new-generation polarimeters. In particular, we take a step 
beyond previous treatments of RCS in magnetar magnetospheres 
by considering different surface emission models which are 
more physically motivated, a magnetized atmosphere and a 
condensed surface, in place of simple blackbody emission. 
Spectro-polarimetric calculations are finalized to produce 
synthetic data for {\it IXPE} in order to evaluate the detectability 
of magnetar sources and the capability of the instrument 
in disentangling different physical situations.

The plan of the paper is as folllows. The theoretical framework
is set up in section \ref{section:themodel}, while section
\ref{section:numericalimplementation} contains a description
of the numerical techniques used to compute spectra and polarization 
observables. Results are presented in
section \ref{section:results} and simulations of {\it IXPE}
observations are reported in section \ref{section:ixpe}.
Discussion and conclusions follow in section \ref{section:discussion}.

\section{The model}
\label{section:themodel}
The spectral properties of the persistent emission from
magnetar sources have been successfully modelled in
terms of the RCS paradigm,
as originally suggested by \citet{tlk02}. According to
this scenario, thermal photons emitted by the cooling
star surface are resonantly up-scattered by charges
flowing along the closed field lines of a non-potential
(twisted) magnetic field. This is expected to produce
a thermal spectrum with an extended power-law like tail
in the $0.1$--$10$ keV energy range. Detailed calculations,
based on Montecarlo simulations, confirmed this picture
(Fern\`{a}ndez \& Thompson 2007; Nobili, Turolla \& Zane
2008), and their application to spectral fitting of
SGRs/AXPs allowed to estimate the physical parameters
of magnetar magnetospheres \cite[][]{zane+09}. Following
the same approach, the expected polarization pattern in
magnetars has been also investigated in the wake of the
growing interest in X-ray polarimetry \cite[][]{fd11,tav+14,guv+15}.

Photons propagating in a strongly magnetized vacuum are
expected to be elliptically polarized in two normal modes,
the ordinary (O) and the extraordinary (X) ones, with
the photon electric field mainly oscillating either parallel
or perpendicular to the plane of the photon propagation
direction and the local magnetic field, respectively
\cite[][see also \citealt{hl06}]{gp73}. Here we assume
that radiation is linearly polarized in the two modes.The
observed polarization pattern depends on the intrinsic 
polarization of the surface thermal radiation, on scatterings 
in the magnetosphere and on QED effects \cite[vacuum
birefringence, see][]{hs00,hs02,fd11,tav+14,tav+15,gonz+16}. Previous
investigations of photon reprocessing in the magnetar 
magnetosphere adopted the simplifying assumption of
100\% polarized (either in the X or O mode) blackbody
radiation for surface emission\footnote{The predominance
of extraordinary photons can be justified on the basis
of the reduced opacity for the X-mode \cite[see e.g.][see
also \citealt{lh02}]{ps78,mesz92}.}. Hence, more realistic
models for surface emission should be considered to
better characterize the spectral and polarization
properties of magnetar persistent emission within the
RCS scenario, as discussed below.

\subsection{Thermal emission from NS surface}
\label{subsection:thermalemission}
\begin{figure}
\begin{center}
\includegraphics[width=7.cm]{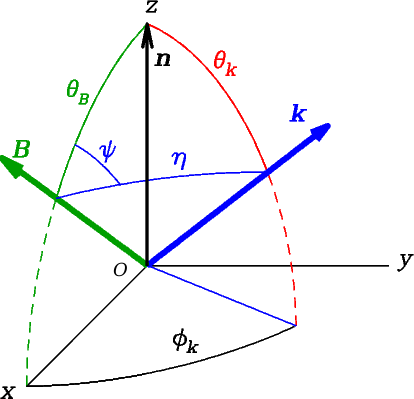}
\caption{Emission geometry for a small element 
on the NS surface centered in $O$. The $z$-axis is 
along the surface normal $\boldsymbol n$, while $\boldsymbol 
B$ and $\boldsymbol k$ are the local magnetic field 
and the unit vector along the line-of-sight, respectively. 
The direction of $\boldsymbol k$ in the $xyz$ frame is 
fixed by the pair of polar angles $\thetak$ and $\phik$. 
The angles $\eta$ and $\psi$ provide again the direction 
of $\boldsymbol k$, but this time with respect to $\boldsymbol B$. 
Finally, $\thetaB$ gives the inclination of $\boldsymbol 
B$ with respect to the surface normal.}
\label{figure:angles}
\end{center}
\end{figure}
The properties of the outermost layers of a highly-magnetized
NS are still debated. According to the commonly accepted
picture, NSs should be covered by a gaseous atmosphere,
with typical scale-height $H\sim kT/m_{\rm p}g\approx
0.1$--$10$ cm, with $T$ the surface temperature, $m_{\rm
p}$ the proton mass and $g$ the surface gravity. Model
atmospheres have been investigated by many authors, allowing
for different chemical compositions (H, He and heavy elements)
and magnetization \cite[see e.g.][for a recent review]{rom87,
shib+92,pav+94,pot14}. On the other hand, magnetic fields
strong enough that the electron gyroradius becomes smaller
than the Bohr radius ($B\ga 2.4\times10^9$ G) change
substantially the properties of matter. Under such
conditions, highly elongated atoms can form molecular
chains and this can result in a phase transition which
turns the star surface into a condensate \cite[either
liquid or solid, see][]{ls97}. The critical temperature
$T_{\rm crit}$ for condensation depends on the magnetic
field and on the composition \cite[][]{lai01,ml07}. In
magnetar sources, where $B\ga10^{14}$ G, magnetic condensation
is expected to take place also at temperatures in excess
of $0.1$ keV for heavy element compositions, in the same range of those inferred
from X-ray observations. Indeed, as it can be seen from Figure \ref{figure:tcrit}, 
some magnetars lie within the condensation region, while 
others, the ``low-field'' SGR 0418+5729 in particular \cite[see][]{rea+10}, 
do not. We note, however, that the values of $B$ and $T$ used to mark
the position of magnetar sources in the plot are taken from 
the McGill catalogue \cite[][]{ok14}. This means that $T$ is 
the blackbody temperature (as obtained from the spectral fit) 
and $B$ is the dipole field (as derived from the spin-down measure
at the magnetic equator). 
Actually, estimates of the emitting area indicate that thermal 
photons come from a heated cap, while the rest of the surface is 
cooler \cite[typically $\approx 1$ vs. $\approx 0.1$ keV; see 
e.g. the discussion in][for the AXPs XTE J1810-197 and CXOU 
J164710.2-455216]{albano+10}. In this respect, the temperatures 
reported in Figure \ref{figure:tcrit} are likely an overestimate, since 
the (single) blackbody fit samples the hottest component most. 
It is also possible that surface condensation depends on the 
(magnetic) co-latitude, with the cooler equatorial belt undergoing 
a phase transition and the hotter caps not. The local, surface 
strength of the $B$-field, which is not likely to coincide with 
the spin-down value, is also key in producing the condensation.

Similar considerations have been already presented 
in analyzing the polarization properties of thermal emission 
from X-ray Dim Isolated Neutron Stars (XDINSs) and magnetars 
by \citet[see also \citealt{sant+19a}]{gonz+16,gonz+17},
who also explored a gaseous atmosphere and a condensed surface.
Magnetar RCS spectral and polarimetric models were 
investigated in the idealized case of BB seed photons 
\cite[][]{ft07,ntz08,zane+09,fd11,tav+14}. An exception
is the work by \citet{guv+15}, in which modelling
of the spectral properties of the AXP 1E 1048.1--5937
was carried out assuming that thermal surface radiation
comes from a magnetized, fully-ionized H atmosphere.
In this investigation we consider surface emission
from both a magnetized H atmosphere and a condensed
surface. The main features of the models we use are
discussed below. In the following all calculations
refer to a NS with mass $M_{\rm NS}=1.5\,M_\odot$ and 
radius $R_{\rm NS}=12$ km (which correspond to a surface 
gravity $g_{\rm NS}=1.4\times10^{14}$ cm s$^{-2}$).




\subsubsection{Magnetized atmosphere}
\begin{figure*}
\begin{center}
\includegraphics[width=17.5cm]{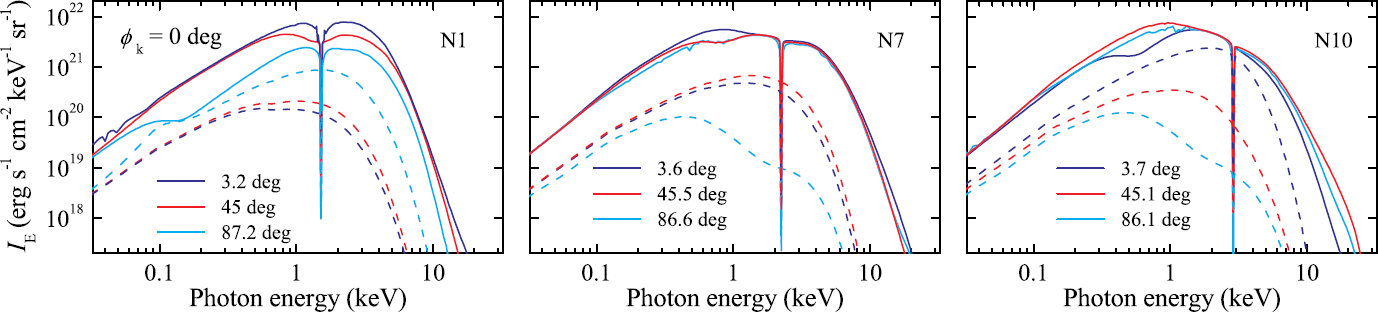}
\caption{Spectral distributions of the specific intensities
$I_{\rm X}$ (solid lines) and $I_{\rm O}$ (dashed lines) for
different values of $\thetak$ (marked on each panel) and 
$\phik=0^\circ$, for the models $1$ (left), $7$ (center)
and $10$ (right, see Table \ref{table:anglesmodB}). Here
the effective temperature is $T_{\rm eff}=0.5$ keV.}
\label{figure:intensityoutput}
\end{center}
\end{figure*}
Highly magnetized plane-parallel NS atmospheres are 
computed using the code described in \citet{sul+09},
suitably modified to account for different inclinations
$\thetaB$ of the magnetic field with respect to the 
local surface normal. Only a pure-hydrogen composition 
is considered and partial ionization in strong magnetic 
fields is properly treated \cite[see e.g.][]{pot+04}.
Vacuum polarization effects (the Mikheyev-Smirnov-Wolfenstein 
resonance) are included and mode switching is treated
following \citet{vanadlai06}.
The opacities, polarizabilities, and equation of state 
of the hydrogen plasma were computed, using the updated 
approach discussed in \citet{pot+14}, on a fixed grid 
of plasma temperature and density, 
from which the values required during the actual calculation 
were obtained by interpolation. For the sake of comparison
with previous works \cite[][]{ntz08,tav+14} and also because 
of the lack of a more accurate model, the surface temperature 
is assumed to be constant (a more thorough discussion on 
the limitations and the effects of this assumption on our 
results is presented in Section \ref{section:discussion}). 
The output (i.e. the specific intensities $I_{\rm X}$ and 
$I_{\rm O}$ for X- and O-mode photons, respectively) was 
produced for a selected sample of magnetic field strengths 
and inclinations (see Table \ref{table:anglesmodB}). 

\begin{table}
\caption{Values of the magnetic colatitude $\Theta_{\rm s}$,
magnetic field strength $B$ and inclination $\thetaB$ used 
for the magnetized atmosphere models (here $\Bp=5\times10^{14}$
G).
\label{table:anglesmodB} }
\begin{center} 
\begin{tabular}{ c |  c c c}
\hline
\hline
 N  & $\cos\Theta_{\rm s}$ & $B/10^{14}$\,G & $\cos \theta_B$ \\ \hline
       1  & 0.05 &   2.40809 &   0.101399 \\
       2  & 0.15 &   2.47787 &   0.295411 \\
       3  & 0.25 &   2.61143 &   0.466226 \\
       4  & 0.35 &   2.79889 &   0.607170 \\
       5  & 0.45 &   3.02928 &   0.718498 \\
       6  & 0.55 &   3.29254 &   0.804283 \\
       7  & 0.65 &   3.58034 &   0.869697 \\
       8  & 0.75 &   3.88628 &   0.919553 \\
       9  & 0.85 &   4.20559 &   0.957813 \\
      10  & 0.95 &   4.53447 &   0.987560 \\
\hline
\hline
\end{tabular}
\end{center}
\end{table}
Model atmospheres for an inclined magnetic field require
much longer computational times with respect to the aligned 
case, essentially because the transfer problem must be 
now solved in two dimensions instead of one, due to the loss 
of symmetry around the surface normal (i.e. the specific intensity 
depends now on both the polar angles and not on the co-latitude alone; 
see below). The optical properties of the magnetized plasma depend on 
the angle $\eta$ between the photon propagation direction 
$\boldsymbol{k}$ and the local magnetic field $\boldsymbol{B}$. 
On the other hand, under the plane-parallel approximation 
the radiation field naturally depends on the angle $\thetak$ 
between $\boldsymbol{k}$ and the surface normal $\boldsymbol{n}$ 
(see Figure \ref{figure:angles}). In the aligned case ($\thetaB
=0$), $\eta$ and $\thetak$ coincide; this allows to solve 
the radiative transfer equation over a one-dimensional 
$\thetak$ angular grid, exploting the axial symmetry
of the specific intensity around $\boldsymbol{B}\equiv
\boldsymbol{n}$. On the other hand, for $\thetaB\ne0$
the symmetry is lost and the transfer
equation should be solved on a two-dimensional ($\thetak,
\phik$) grid, with $\phik$ the azimuth associated to
$\thetak$. To avoid the interpolation of the opacities 
over such a two-dimensional grid, the code solves the transfer 
problem over an ($\eta,\psi$) angular grid, with $\psi$ 
the azimuth associated to $\eta$ (see Figure \ref{figure:angles}). 
However, in order to compute integrals over the NS surface one needs anyway 
to know the values of $\thetak$ and $\phik$ corresponding 
to each point in the ($\eta,\psi$) grid. These can be obtained 
through a spherical polar coordinate transformation,
\begin{flalign} \label{equation:etapsitothetaphi}
\cos\thetak &= \cos\thetaB\cos\eta+\sin\thetaB\sin\eta\cos\psi \nonumber & \\
\cos\phik &= \frac{\sin\thetaB\cos\eta-\cos\thetaB\sin\eta\cos\psi}{\sin\thetak}\nonumber & \\
\sin\phik &= \frac{\sin\eta\sin\psi}{\sin\thetak}\,, &
\end{flalign}
where the correct value of $\phi_{\rm k}$ depends on the 
signs of both $\cos\phik$ and $\sin\phik$. Actually, only 
the models corresponding to the northern magnetic hemisphere 
($\cos\thetaB>0$) need to be computed. In the southern hemisphere 
the replacement $\phik\rightarrow\pi-\phik$ holds 
for the patches characterized by $\thetaB>\pi$ because of 
the symmetry properties of the opacities \cite[see e.g.][]{gonz+16}\footnote{As
it will be discussed later on, actual calculations refer to
a globally-twisted dipole field, which however possesses 
the same north-south symmetry as a dipolar field \cite[see][]{tlk02,
pav+09}.}. This relation translates into $\eta\rightarrow
\pi-\eta$ in the coordinate system referred to $\boldsymbol{B}$,
as it can be checked using equations (\ref{equation:etapsitothetaphi}).
The specific intensity emerging from the atmosphere is shown 
in Figure \ref{figure:intensityoutput} for three models with different 
magnetic field strength and inclination, and $T_{\mathrm{eff}}=0.5$ 
keV (\# 1, 7 and 10 in table \ref{table:anglesmodB}); results refer 
to $\phik=0$ and different values of $\thetak$. A distinctive feature 
is the presence of a narrow proton cycloton line in absorption at 
$E_\mathrm{Bp}\sim 0.63(B/10^{14}\, \mathrm{G})$; the high degree 
of anisotropy in the emission is clearly seen by comparing the different 
curves in each panel.

In a twisted magnetosphere charges flow along the close 
magnetic field lines. This implies that returning currents 
hit the star surface, heating the star external layers 
\cite[see][]{bt07,fd11}. Investigations on the effects 
of backflowing charges on the atmospheres have started 
\cite[][]{gonz+19}, but a complete study is still 
lacking. For this reason, in the present work we consider 
a passively cooling, magnetized atmosphere, neglecting all 
the possible effects of returning currents (see again 
Section \ref{section:discussion} for a more comprehensive 
discussion).

\subsubsection{Condensed surface}

Thermal radiation coming from a bare NS was first studied 
by \citet{brink80} and further investigated by \citet[see 
also \citealt{pot14} for a recent review]{tzd04,paz+05,vanad+05,pot+12}. 
In these papers, the calculation of the intensity and
polarization of radiation emitted by a condensed surface was 
based on the (complex) dielectric tensor in the medium just
below the condensed surface, derived in the cold plasma 
approximation \cite[see e.g.][]{gin70}. It should be noted that this formalism
allows to treat not only reflection and transmission of an
electromagnetic wave, but also true absorption, which enters 
the equations through the imaginary part of the indices of refraction. 
The monochromatic intensities of the two independent polarization modes 
$\ell=1,2$ emitted by a heated condensed surface\footnote{Modes $1$ and $2$ 
are defined as perpendicular and parallel, respectively, to the plane of 
the photon direction $\kvec$ and the local surface normal 
$\nvec$.} can be written as \cite[see][]{vanad+05}
\begin{flalign}
I_{\nu,{\ell}}&= j_{\nu, \ell}\,B_\nu / 2 = (1-R_{\nu, \ell})\,B_\nu / 2, &
\end{flalign}
where $B_\nu$ is the Planck spectral radiance, $j_{\nu, \ell}$ are 
the (normalized) emissivities and $R_{\nu,\ell}$ the reflectivities. 
The latter can be derived by considering 
the reflected and transmitted electromagnetic fields 
($\boldsymbol{E}^{(\rm r)}_{\ell}$ and $\boldsymbol{E}^{(\rm t)}_{\ell}$) 
that arise in response to the incidence of an arbitrary linearly 
polarized wave on the surface, with electric field $\boldsymbol{E}_\ell^{\mathrm{(i)}}=
\mathcal{A}_\ell\boldsymbol{e}_\ell^{\mathrm{(i)}}$,
\begin{flalign}
\boldsymbol{E}^{({\rm r})}_{\ell} &= \mathcal{A}_{\ell} \sum_{{m}=1}^2
r_{m\ell} \,\boldsymbol{e}^{({\rm r})}_{m},
\qquad
\boldsymbol{E}^{({\rm t})}_{\ell} = \mathcal{A}_{\ell} \sum_{{m}=1}^2
t_{m\ell} \,\boldsymbol{e}^{({\rm t})}_{m}\,; &
\end{flalign}
here $\mathcal{A}_{\ell}$ are the complex amplitudes of the electric 
field of the incoming wave, $\boldsymbol{e}_\ell^{\mathrm{(i)}}$, 
$\boldsymbol{e}_{\ell}^{\mathrm{(r)}}$ and 
$\boldsymbol{e}_{\ell}^{\mathrm{(t)}}$ are the unit polarization vectors 
and $r_{m\ell}$ and $t_{m\ell}$ are complex coefficients.
The dimensionless reflectivities for the two orthogonal linear
polarizations can be expressed as
\begin{flalign}
R_{\nu,\ell} &= |r_{\ell1}|^2+|r_{\ell2}|^2\,. &
\end{flalign}
The coefficients $r_{m\ell}$ are in turn obtained
from Maxwell's equations for the transmitted modes,
\begin{flalign}
\left[\boldsymbol{\epsilon} + n^2_{\ell}(\boldsymbol{k}^{({\rm t})}_{\ell}
\otimes\boldsymbol{k}^{({\rm t})}_{\ell} - \mathbb{1})\right]\cdot 
\boldsymbol{E}^{({\rm t})}_{\ell} &= \mathbf{0}\,, &
\end{flalign}
where $\boldsymbol{\epsilon}$ is the dielectric tensor of the medium,
$n_{\ell}$ is the index of refraction for mode $\ell$, $\boldsymbol{k}
^{({\rm t})}_{\ell}$ is the (transmitted) unit wavevector 
and $\mathbb{1}$ is the unit tensor. Maxwell's boundary conditions
connect $r_{m\ell}$ with $t_{m\ell}$, thus closing this set of equations.
 
The exact form of $\boldsymbol{\epsilon}$ is currently unknown. One
usually applies the approximation of a fully ionized electron-ion plasma
with either non-moving ions (the so-called fixed-ion limit) or
finite-mass ions freely responsive to the electromagnetic forces (free-ion
limit). In reality, however, there are Coulomb forces between the ions, which
hamper the ion response, but still can not completely ``freeze'' the
ions in their equilibrium positions. The problem is further complicated
by the possibility of incomplete ionization, i.e. the existence of bound
electron shells embedded in the plasma, which may be anticipated in a
shallow layer beneath the surface.

In the following we use the analytical, approximating formulae
developed by \citet[see also \citealt{gonz+16}]{pot+12}, which
provide the emissivities $j_{\nu,\ell}$
as functions of the magnetic field strength $B$, the photon
energy $h\nu$, the angle $\thetaB$ and the polar angles $\thetak$
and $\phik$ between the local normal $\nvec$ to the star surface
and the photon direction $\kvec$. The expressions for the emissivities
$j_{\nu,\rm X}$ and $j_{\nu,\rm O}$ in the X and O modes are then
obtained from $j_{\nu,1}$ and $j_{\nu,2}$ through a rotation
in the plane orthogonal to $\kvec$, which amounts to write
\begin{flalign} \label{equation:rotation12XO}
j_{\nu,\rm O} &= \left(\boldsymbol{x}'_{\rm r}\cdot\boldsymbol{e}_1^{(\rm r)}\right)^2j_{\nu,1} +
                 \left(\boldsymbol{x}'_{\rm r}\cdot\boldsymbol{e}_2^{(\rm r)}\right)^2j_{\nu,2} \nonumber & \\
j_{\nu,\rm X} &= \left(\boldsymbol{y}'_{\rm r}\cdot\boldsymbol{e}_1^{(\rm r)}\right)^2j_{\nu,1} +
                 \left(\boldsymbol{y}'_{\rm r}\cdot\boldsymbol{e}_2^{(\rm r)}\right)^2j_{\nu,2}\,, &
\end{flalign}
where the quantities $\boldsymbol{x}'_{\rm r}\cdot\boldsymbol{e}_{\ell}
^{(\rm r)}$ and $\boldsymbol{y}'_{\rm r}\cdot\boldsymbol{e}_{\ell}
^{(\rm r)}$ are given in Appendix B of \citet{pot+12}\footnote{Note 
that a typo is present in equation (B.12) of \citet{pot+12}, where 
$\cos^2\thetak-\sin^2\thetak$ should be $\cos^2\thetak+\sin^2\thetak$. 
The correct expression can be found in arXiv:1208.6582.}. 
We note that equations (\ref{equation:rotation12XO}) hold if $j_{\nu,1}$ 
and $j_{\nu,2}$ are mutually incoherent. Actually, under the conditions 
considered in this work, the differences introduced by applying the complete 
transformation of \citet{pot+12} are negligible 
(except possibly at $10$ keV), as we checked numerically.
The corresponding intensities are then expressed as
\begin{flalign} \label{equation:bsintens}
I_{\nu,\rm O} &= j_{\nu,\rm O}(B,\nu,\thetaB,\thetak,\phik)B_\nu(T)/2 \nonumber & \\
I_{\nu,\rm X} &= j_{\nu,\rm X}(B,\nu,\thetaB,\thetak,\phik)B_\nu(T)/2\,, &
\end{flalign}
Following \citet[see also \citealt{pot+12}]{vanad+05}, our calculations are performed
in the simplifying limits of free and fixed ions. As already noted by \citet{vanad+05},
a more realistic description of the actual reflectivity
can be expected to lie in between the previous limits.
Finally we warn that, as for the atmosphere model discussed
above, also in this case no allowance for the effects of
returning currents is made.


\section{Numerical implementation}
\label{section:numericalimplementation}
The present investigation relies on the Montecarlo code
developed by \citet{ntz08} for simulating the persistent
emission from magnetar sources in the framework of the
RCS model. The polarization observables (linear polarization
degree and angle) are computed accounting for the evolution
of polarization states in the magnetized vacuum (vacuum
birefringence) by a specific module \cite[see][for further
details]{tav+14}. We assumed a globally twisted dipole field,
characterized by the polar strength $\Bp$ and the radial
index $p$,
\begin{flalign} \label{equation:Btwisted}
\boldsymbol{B} & = \left(B_r,B_\theta,B_\phi\right) & \nonumber \\
& = \frac{B_{\mathrm{p}}}{2} \left(\frac{r}{R_{\mathrm{NS}}}\right)^{-p-2}
\left[-f', \frac{pf}{\sin\theta},
\sqrt{\frac{C(p)\ p}{p+1}} \frac{f^{1+1/p}}{\sin\theta}\right]\,, &
\end{flalign}
where the angular part of the flux function $f(\mu)$
satisfies the Grad-Schl\"{u}ter-Shafranov equation
\begin{flalign} \label{equation:GSS}
(1-\mu^2)f''+p(p+1)f+Cf^{1+2/p}&=0\,; &
\end{flalign}
here $\mu=\cos\theta$, a prime denotes the derivative
wrt $\mu$ and the constant $C$ is an eigenvalue \cite[][]{tlk02,
pav+09}. We remark that globally twisted dipole fields are 
a particular case of the much more general force-free equilibria, 
obtained  by adding a defined amount of shear to the (potential) 
dipole field. The shear itself is expressed in terms of $p$ or, more
conveniently, through the twist angle $\Deltaphins$,
\begin{flalign} \label{twistang}
\Delta \phi_{\rm N-S} &= \lim_{\theta_0 \to 0} 2\int_{\theta_0}^{\pi/2} \frac{B_\phi
}{B_\theta } \frac {d\theta}{\sin \theta}\,, & 
\end{flalign}
which measures the amount of angular displacement suffered 
by a field line in going from the north to the south magnetic pole.

Charges are assumed to flow along the closed field lines with 
constant velocity $\beta$ (in units of the speed of light). We retain
the original hypothesis introduced in \citet[see also \citealt{ntz08}]{ft07}
of an unidirectional flow of electrons, streaming from the north to
the south magnetic hemisphere (see section \ref{section:discussion}
for further details).

Emission from the star is handled by dividing the
surface into a number of equal-area patches by means of a
$\Theta_{\rm s}$--$\Phi_{\rm s}$ grid, where $\Theta_{\rm
s}$ is the magnetic colatitude and $\Phi_{\rm s}$ the azimuth
of the patch center. Photons are eventually collected onto
a sphere representing the observer's sky \cite[see][for
further details]{ntz08}. In the actual calculations a
$10\times10$ and a $20\times10$ angular meshes were used
in the case of condensed surface and atmosphere models\footnote{The
latter is given in Table \ref{table:anglesmodB}
for the northern hemisphere (the pathces in the southern
hemisphere are obtained by mirroring with respect
to the magnetic equator).},
respectively, while a $15\times15$ grid was adopted for 
the sky-at-infinity in all cases. The number of seed photons 
is arbitrarily fixed for a reference patch and follows 
from the scaling with the number flux for the remaining 
ones (e.g. $T^3$ in the case of blackbody emission)\footnote{The 
reference patch is actually chosen as the one which emits 
the lowest number of photons.}. In the original version 
of the code, a blackbody photon distribution at the local 
temperature $T$ was assumed, together with the linear polarization
fraction (either $100\%$ polarized in the X- or O-mode,
or unpolarized, i.e. $50\%$ X and $50\%$ O). Now instead,
the polarization degree of the emitted radiation consistently
follows once the surface emission model is specified.
No allowance for general relativistic 
(GR) effects is made in the calculations, mostly to avoid 
an undue increase in computational time when running a large 
number of Montecarlo simulations. This is tantamount to ignore 
light bending and gravitational redshift. GR light bending 
outside a NS (assuming a Schwarzschild space-time) exposes 
a larger part of the star surface to a far away observer 
(actually, for typical values of the NS mass and radius, 
more than $2/3$ of the entire surface will be visible at 
the same time). This deeply impacts e.g. on the observed 
pulsed fraction for surface emission from cooling, isolated 
NSs \cite[see e.g.][]{belo02,tn13}. In the case of magnetar 
sources, where RCS is at work, the effects are likely less 
important, because a sizable part of the radiation emitted 
by the surface will scatter at $\sim 5$--$10$ star radii, 
where GR effects already abated, before reaching infinity. 
Results obtained with Montecarlo codes including or not GR 
ray-tracing are, in fact, in good agreement \cite[][]{fd11,tav+14}.
Strong gravity can also influence the photon polarization 
state, because of the rotation of the polarization plane
along the null geodesics \cite[][]{cs77,sc77,cps80}. However, 
the typical scale-length over which vacuum birefringence 
acts is much shorter than that of gravity, so that the effects
of the latter on polarization are negligible around ultra-magnetized 
NSs.

\subsection{Sampling the seed photon distribution}
The starting point of the Montecarlo simulation is to
generate seed photons randomly distributed according to
the specific intensity of the radiation emitted from the
star surface. Obtaining a random deviate from a Planckian
distribution poses in general no difficulties on a numerical
ground, since an efficient method can be devised despite
the fact that the associated cumulative density function
(cdf) is not analytically invertible. On the other hand,
generating a random deviate in the case of emission from
a condensed surface or a magnetized atmosphere requires
a greater computational effort. A simple way of dealing
with this is using the acceptance/rejection method
\cite[][]{vN51,numrec}, which can be exploited in those
cases in which a random variable has to be generated from
a probability density function (pdf) $f$ with non-invertible
(or not even computable) cdf. The key point is to find
a function $g$ (the majorizing function) such that $g>f$
in the desired range and from which a random deviate
can be easily obtained. Once a random variable $x$ is
extracted from the normalized majorizing distribution
$\bar{g}$, the ratio $h(x)=f(x)/g(x)$ is calculated
and, if the condition $y<h(x)$ (with $y$ an uniform
random deviate between $0$ and $1$) is met, then $x$
provides a sampling of the starting pdf $f$.

In the case at hand, the anisotropic emission of both 
the atmosphere and the condensed surface requires an 
extension of the acceptance/rejection technique to 
three dimensions: the photon frequency $\nu$ and the 
two angles $\thetak$ and $\phik$ (or $\eta$ and $\psi$) 
which fix the photon direction. This can be addressed 
by choosing a majorizing function $g_{\nu,j}$ independent 
of the angles and such that $g_{\nu,j}>I_{\nu,j}(\thetak,
\phik)$ for any value of $\thetak$ and $\phik$ ($j={\rm O, X}$). 
In this way the photon energy is generated 
according to the distribution $I_{\nu,j}$, while the two 
angles are extracted from uniform deviates, in the ranges $0
\leq\eta\leq\pi$ and $0\leq\psi\leq\pi$ for the atmosphere 
model, and $0\leq\thetak\leq\pi/2$ and $0\leq\phik\leq2\pi$ 
for the condensed surface one\footnote{In the atmosphere case, 
some values of $(\eta,\psi)$ may actually label incoming 
rays, i.e. the corresponding value of $\thetak$, as derived 
from the first of equations (\ref{equation:etapsitothetaphi}), 
turns out to be $>\pi/2$. When this happens, symmetry considerations
allow to associate the intensity $I_\nu(\eta,\psi)$
to the direction characterized by the angles $\pi-
\thetak$ and $\pi+\phik$.}. Care must be taken that $g_{\nu,j}$ is
as close as possible to $I_{\nu,j}$, in order
for the rejection method to be efficient. We found
that a Planckian shape for $g_{\nu,j}$ with a suitable
scaling works well in all cases. This choice has also
the advantage that our Montecarlo code already contains
an algorithm to generate random deviates from a normalized
Planckian distribution. This ensures that photons are emitted with the
correct angular distribution, as we numerically tested. 
For the atmosphere model, in which the intensities are 
tabulated, a three-dimensional, linear interpolation 
is performed in order to evaluate $I_\nu$ at the required
photon energies and angles.

\subsection{Polarization mode evolution}

In order to determine the initial polarization state
of each emitted photon, the code numerically integrates
the specific intensity to get, for each surface patch, 
the ratio $J_{\rm X}/J$, where $J$ ($J_{\rm X}$) is the
gray total (X-mode) mean intensity. For each emitted photon an
uniform random deviate $U$ is then generated: the
photon is labelled as extraordinary if $U<J_{\rm X}/J$
and as ordinary otherwise.

Since the typical densities in magnetar magnetospheres 
are $\approx 10^{16}$ cm$^{-3}$, polarization transport 
is dominated by the dielectric and magnetic permeability 
properties of the magnetized vacuum, as it follows from 
the equation of plane waves \cite[see e.g.][and references 
therein]{hl06}. Closer to the star surface, where the magnetic
field is stronger, the effect of magnetized vacuum is to 
lock the photon polarization vector to the local magnetic
field direction (adiabatic region). As radiation propagates 
outwards, the polarization vector progressively decouples 
(intermediate region) until, at large distances, it freezes 
(external region). Since under typical magnetar conditions 
RCS occurs well inside the adiabatic region, in the Montecarlo 
code the calculation of the Stokes parameter evolution 
starts once the escape condition is met, i.e. when the 
probability for photons to undergo further resonant scatterings 
has sufficiently dropped \cite[][]{fd11,tav+14}. First the 
Stokes parameters are rotated, so as to express them in a 
(common) fixed frame; then the integration of the evolution 
equations is performed until a large enough radius is reached. 
Some improvements over the original treatment were introduced. 
In particular, care has been taken in selecting the starting 
point for the Stokes parameter evolution always inside the 
adiabatic region, to correctly handle also low energy photons 
which never scatter\footnote{We note that this is not going 
to produce any effect at X-ray energies, while it is potentially 
important in the optical range.}.

\section{Results}
\label{section:results}
In the following we present Montecarlo simulations of 
magnetar persistent emission both in the case of a condensed
surface (in the free/fixed ion limits) and of a magnetized 
hydrogen atmosphere. In particular we highlight the similarities 
and differences with respect to previous results obtained 
for blackbody surface emission 100\% polarized \cite[see][]{ntz08,fd11,tav+14}. 
We remark that this does not correspond to any physical 
situation but provides a simple way to investigate how the 
polarization of (thermal) photons emitted by the surface is 
influenced by magnetospheric RCS and QED effects before reaching 
the observer.
All models have been computed assuming a unidirectional 
(electron) flow, for a polar magnetic field strength $B_{\rm 
p}=5\times10^{14}$ G and a (homogeneous) surface temperature 
$kT=0.5$ keV. These values are representative of magnetar 
sources \cite[][]{ok14,kb17} and close to those inferred 
for the AXPs 1RXS J170849.0$-$400910 \cite[in the following 
J1708, $\Bp=9\times10^{14}$ G, $T=0.5$ keV; see][]{rea+07,dk14} 
and 4U 0142+61 \cite[$\Bp=3\times10^{14}$ G, $T=0.4$ keV; see][]{rea+074u}, 
two selected targets for polarimetric measurements 
with {\it IXPE} (see section \ref{section:ixpe}). The velocity 
spread of magnetospheric electrons is accounted for assuming 
a 1D Maxwellian distribution with temperature $kT_{\rm e}=10$ 
keV. The charge density along the closed field lines is given 
by
\begin{flalign} \label{eledens}
n_\mathrm e & \sim 3\times 10^{15}\left(\frac{B_\phi}{B_\theta}\right)\left(\frac{B}{10^{13}\, \mathrm G}\right)
\left(\frac{r}{10^6\, \mathrm{cm}}\right)^{-1}\vert\langle\beta\rangle\vert^{-1}\ \mathrm{cm}^{-3}\,, &
\end{flalign}
where $r$ is the radial distance and $\langle\beta\rangle$ 
the average charge speed \cite[arising from the superposition 
of bulk and thermal motions; see e.g.][]{tlk02,ntz08}. All 
runs were performed emitting $80,000$ photons from 
the reference patch ($40,000$ for the atmosphere model), so 
that the total number of photons is about $10^7$. 

We produced a set of models varying the two free parameters 
$\Delta\phi_{\rm N-S}$ and $\beta$ in the ranges $[0.3,1.4]$
rad (step $0.1$) and $[0.2,0.7]$ (step $0.1$, with the addition 
of $\beta=0.34$), respectively. Such an archive of models
will be useful especially for polarimetry simulations (see 
section \ref{subsect:polarization}). The source geometry is
characterized by the two angles $\chi$ and $\xi$ that the 
rotation axis makes with the observer's line-of-sight (LOS) 
and the star magnetic axis, respectively. Geometric effects 
are then incorporated at the post-processing level using a 
suite of {\sc idl} codes \cite[see][]{ntz08,tav+14}.

\subsection{Spectra} \label{subsect:spectra}

\begin{figure*}
\begin{center}
\includegraphics[width=17.5cm]{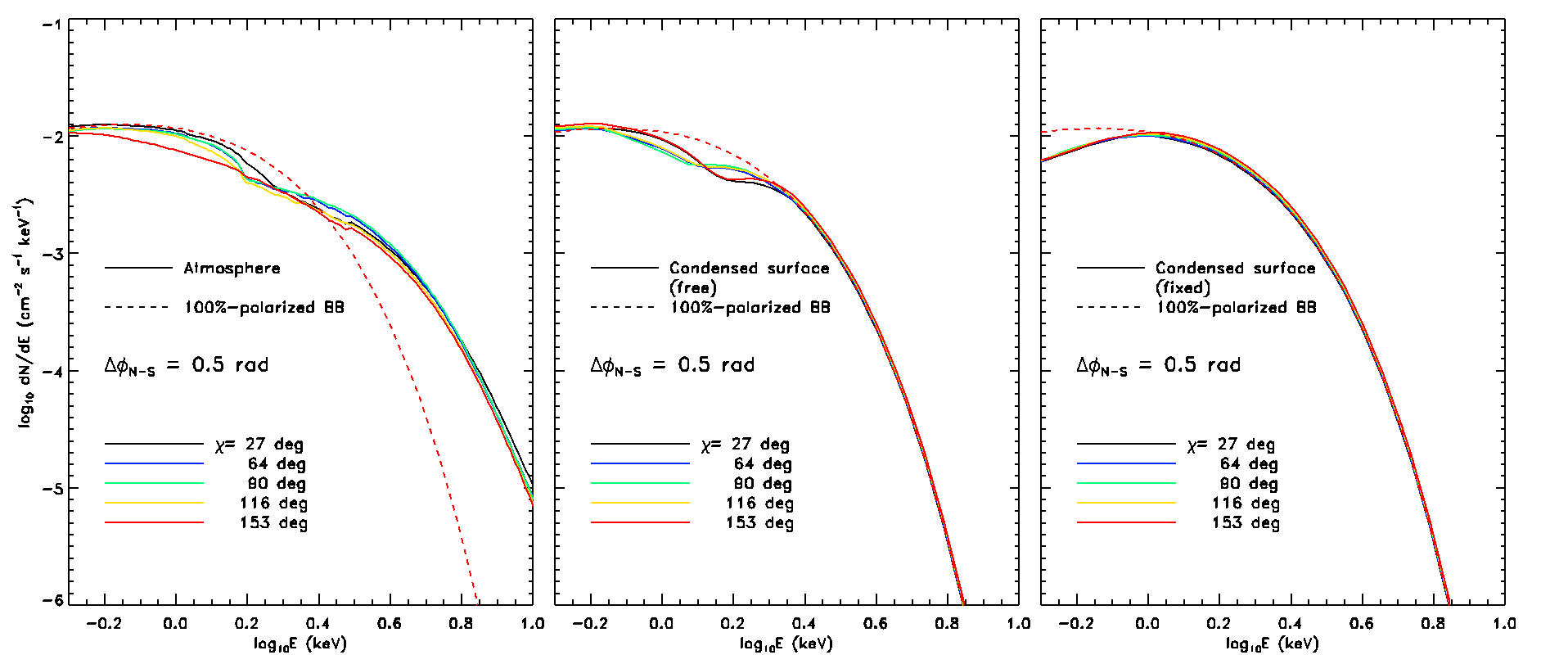}
\caption{Number flux as a function 
of the photon energy for the atmosphere (left) and condensed 
surface models (free-ion middle, fixed-ion right) as seen by 
an observer at infinity. Here the external magnetic field is 
a globally-twisted dipole with $B_{\rm p}=5\times10^{14}$ G 
and $\Deltaphins=0.5$ rad. The star is an aligned rotator 
($\xi=0$) seen at different inclinations $\chi$. The dashed 
lines show the $100\%$-polarized blackbody spectrum for the 
same values of the parameters. Here, no effects from
magnetospheric scatterings are accounted for (see text for 
details).}
\label{figure:specsurf}
\end{center}
\end{figure*}

In order to better understand the effects of RCS on the surface 
emission, we start reporting in Figure \ref{figure:specsurf} the 
spectra of primary photons, i.e. assuming no reprocessing in the 
magnetosphere, as measured by an observer at infinity, accounting 
for emission from the entire star surface for each of the three 
models considered. Spectral (and polarimetric, see Figure
\ref{figure:PiLmorechi}) calculations were performed using 
the ray-tracing code discussed in \citet[see also \citealt{zt06}]{tav+15}.
Here the star is an aligned rotator, and the different solid curves 
refer to different inclinations of the LOS; results for 100\%-polarized, 
blackbody seed photons are also shown for comparison (dashed lines). 
The external magnetic field is a globally-twisted dipole with $\Deltaphins=0.5$ 
rad and $B_{\rm p}=5\times10^{14}$ G.
The atmospheric model shows the characteristic hardening with respect 
to the blackbody at $E\ga kT=0.5$ keV and blending of the proton cyclotron line (which 
occurs at different energies since $B$ changes from patch to patch,
see e.g. Figure \ref{figure:intensityoutput}) produces a broad depression 
around $\sim 2$ keV. Following \citet{ho+08}, to avoid numerical
oscillations in the $\approx 1.5$--$3$ keV range, the specific intensities 
have been re-interpolated over a $E/E_{\rm Bp}$ grid (with $E_{\rm Bp}$
the proton cyclotron energy) before integrating over the visible 
part of the star surface. Condensed surface spectra are closer to the 
blackbody, although an absorption feature appears in the free-ions 
case and the distinct low-energy cut-off is visible in the fixed-ions 
limit.

\begin{figure*}
\begin{center}
\includegraphics[width=17.5cm]{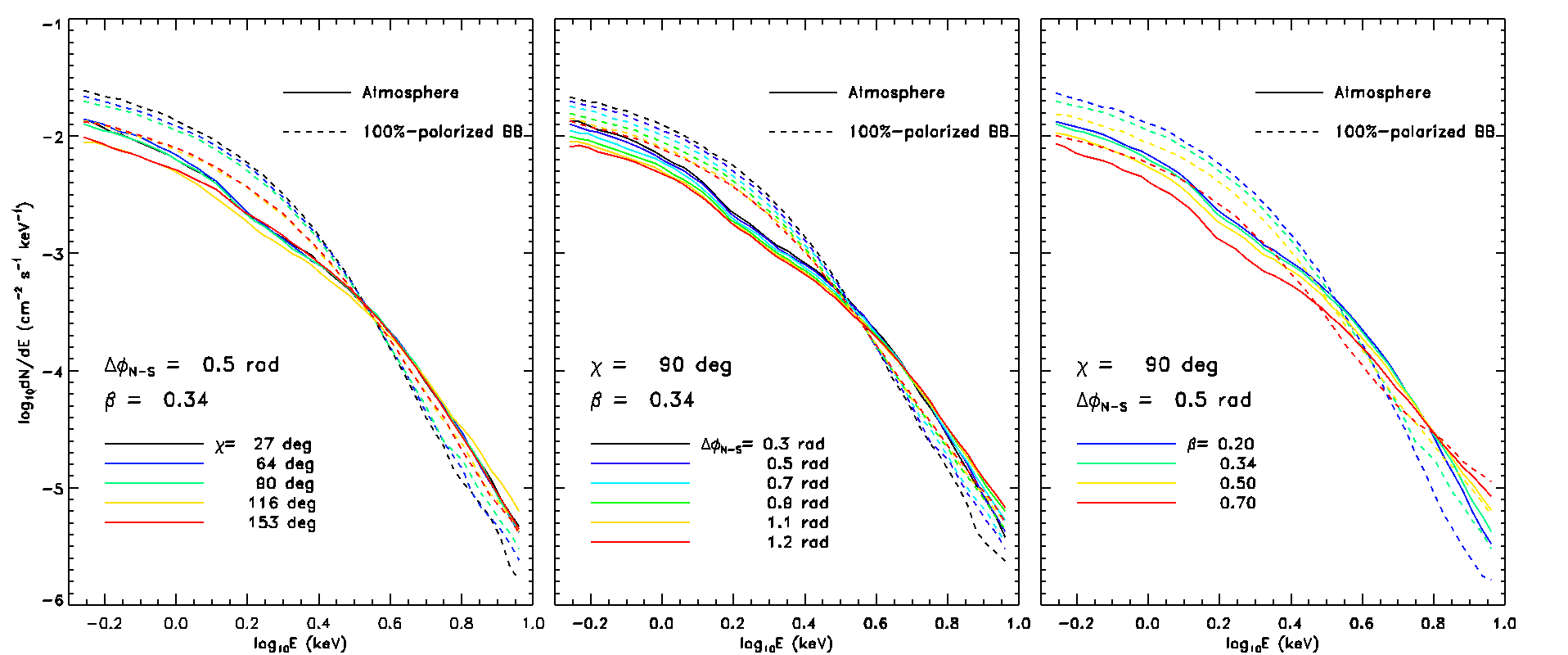}
\caption{Number flux as a function of the photon energy
computed in the case of emission from a magnetized, partially ionized H
atmosphere (solid lines). The plots refer to an aligned rotator
($\xi=0$) with polar magnetic field $B_{\rm p}=5\times10^{14}$ G (see text
for more details). The left, middle and right panels show spectra for
different values of the LOS inclination ($\chi$), the twist angle
($\Deltaphins$) and the electron velocity ($\beta$), respectively.
Spectra for the same values of the parameters but in the case of 100\%
polarized blackbody seed photons are also shown (dashed lines) for comparison.}
\label{figure:specatmo}
\end{center}
\end{figure*}

\begin{figure*}
\begin{center}
\includegraphics[width=17.5cm]{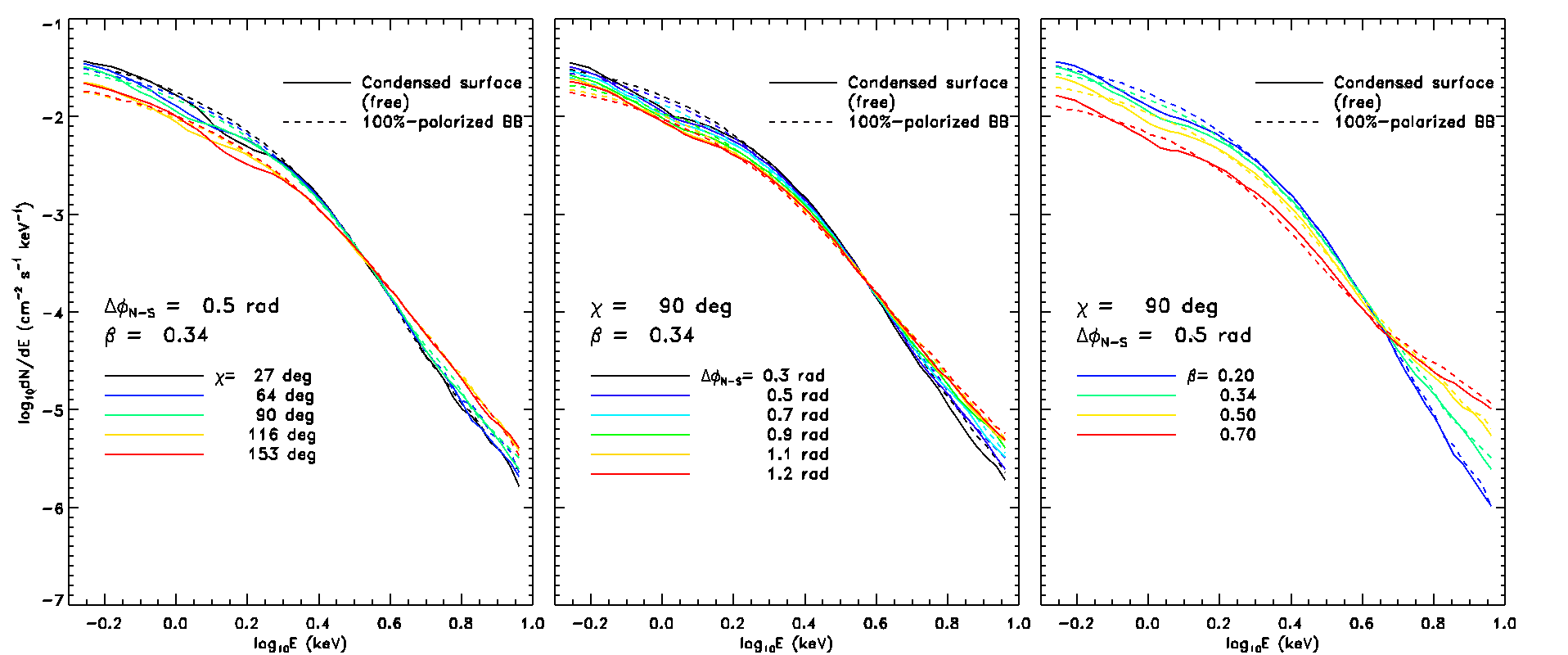}
\caption{Same as in Figure \ref{figure:specatmo} but for condensed surface
emission in the free-ion limit.}
\label{figure:specfree}
\end{center}
\end{figure*}

\begin{figure*}
\begin{center}
\includegraphics[width=17.5cm]{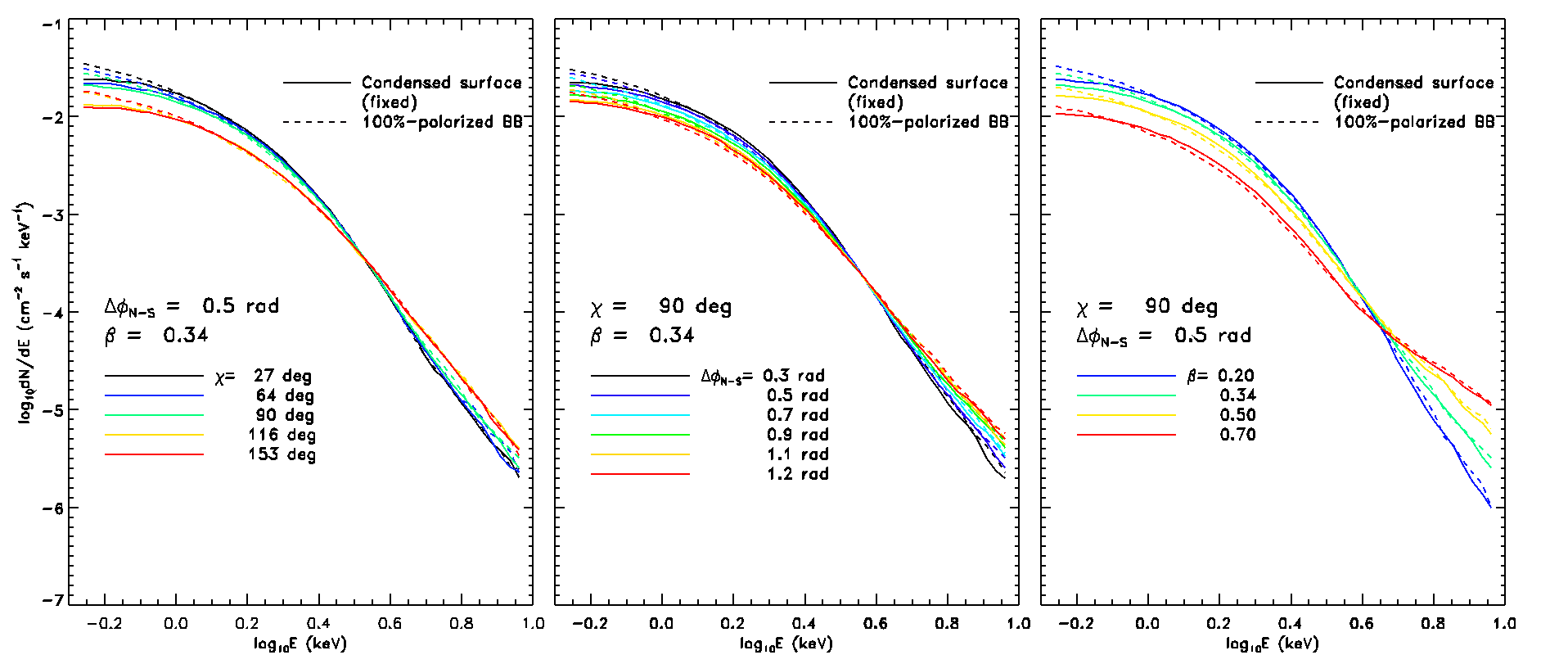}
\caption{Same as in Figure \ref{figure:specatmo}, but for condensed surface
emission in the fixed-ion limit.}
\label{figure:specfix}
\end{center}
\end{figure*}

The number flux as a function of the photon 
energy in the full case in which also RCS is included 
is plotted, instead, in Figure \ref{figure:specatmo} for 
the atmosphere model, and in Figures \ref{figure:specfree} 
and \ref{figure:specfix} for the condensed surface one 
in the free- and fixed-ion limits, respectively. The 
star is still an aligned rotator ($\xi=0$), 
and the different solid curves correspond to different 
values of the LOS inclination $\chi$ (left), twist angle 
$\Delta\phi_{\rm N-S}$ (middle) and electron velocity $\beta$ 
(right). The same spectra but for 100\% polarized blackbody 
seed photons are also shown for comparison (dashed lines). 
In this latter case, as discussed in detail
by \citet{ft07} and \citet{ntz08}, spectra exhibit a distinctive
``blackbody $+$ power-law'' behavior in the $0.5$--$10$ keV
energy range, with the spectral hardness increasing for
increasing values of $\Deltaphins$ and $\beta$, as it can
be clearly seen from the dashed lines of Figures
\ref{figure:specatmo}--\ref{figure:specfix}. When considering
different surface emission models, the overall spectral
shape turns out to still follow a ``thermal $+$ power-law''
behavior. Deviations with respect to the 100\%
polarized blackbody case however appear and are especially 
evident in the low energy range ($0.5$--$2$
keV), where the primary photon distributions peak and hence
a larger fraction of photons reaches infinity without undergoing 
scatterings. In the light of this, such deviations with
respect to the 100\% polarized blackbody
clearly reflect the intrinsic differences among the 
considered surface emission models, that are particularly 
pronounced in the case of atmospheric seed photons (Figure 
\ref{figure:specatmo}), where the imprint of the broad absorption 
line is visible at few keVs. Smaller differences are also present 
in the case of free ions (Figure \ref{figure:specfree}), 
with the appearance of an ``absorption feature'' at around 
$1.2$ keV, intrinsic to the emissivity distribution and related 
to the proton cyclotron resonance and electron plasma frequency 
\cite[see][]{vanad+05,pot+12}.

On the other hand, not unexpectedly, the power-law tails 
are little affected by the details of the seed photon distribution,
so that the spectral indices, which depend only on the 
magnetospheric parameters, turn out to be quite the same 
for all the considered models. However, at high energies 
some differences appear. In fact, for the condensed surface the power-law
tails develop at lower energies ($\sim 3$ keV), much as
in the blackbody case, while for the atmosphere this occurs
only above $\sim 5$ keV, as expected since the seed photon 
distribution is harder (see left panel of Figure \ref{figure:specsurf}).

\subsection{Polarization} \label{subsect:polarization}
\begin{figure*}
\begin{center}
\includegraphics[width=17.5cm]{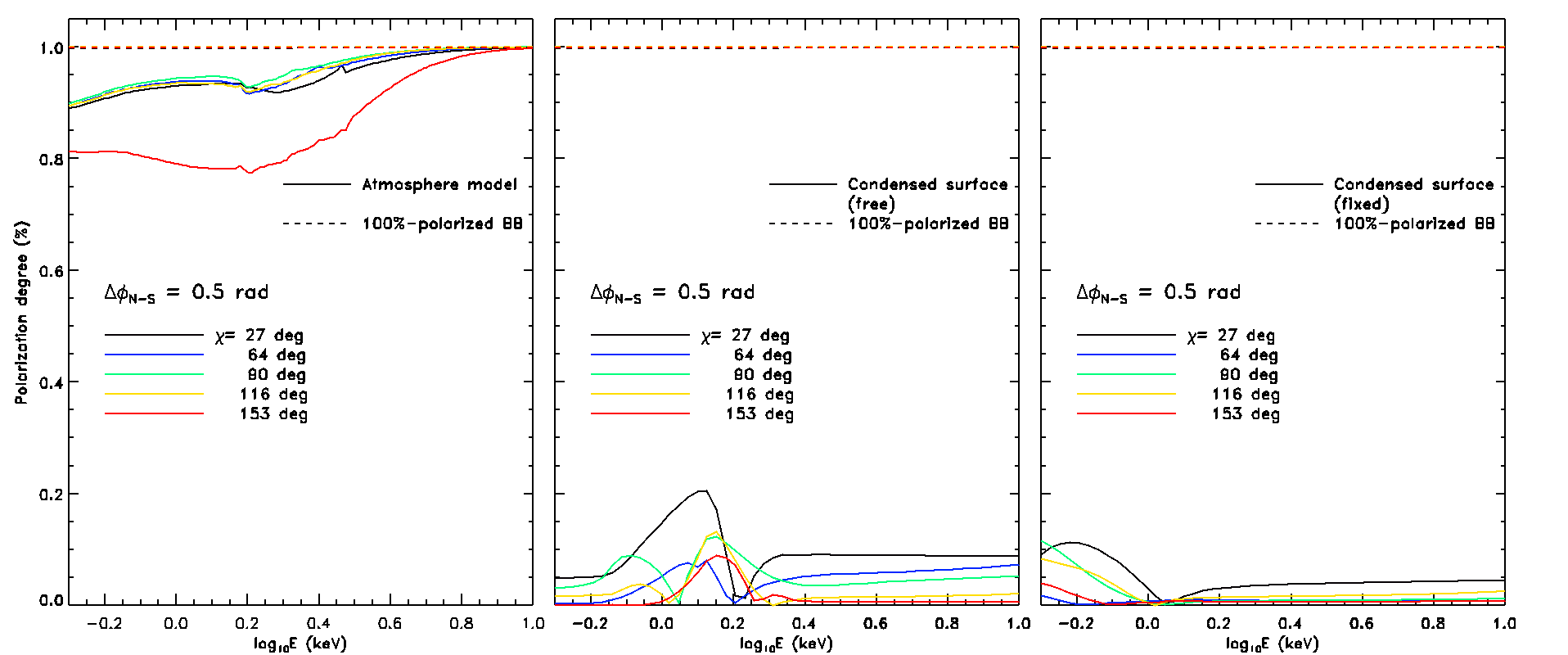}
\caption{Linear polarization degree as a
function of the photon energy of the surface radiation
for the atmosphere (left), condensed surface free-ion
(middle) and fixed-ion (right) models as seen by an observer
at infinity. Here the external magnetic field is a
globally-twisted dipole with $B_{\rm p}=5\times10^{14}$
G and $\Deltaphins=0.5$ rad. The star is an aligned rotator
($\xi=0$) seen at different inclinations $\chi$. The dashed
lines show the intrinsic polarization for the same values
of the parameters for blackbody, $100\%$-polarized seed
photons. Here, no effects from magnetospheric scatterings
are accounted for (see text for details).}
\label{figure:PiLmorechi}
\end{center}
\end{figure*}
\begin{figure*}
\begin{center}
\includegraphics[width=17.5cm]{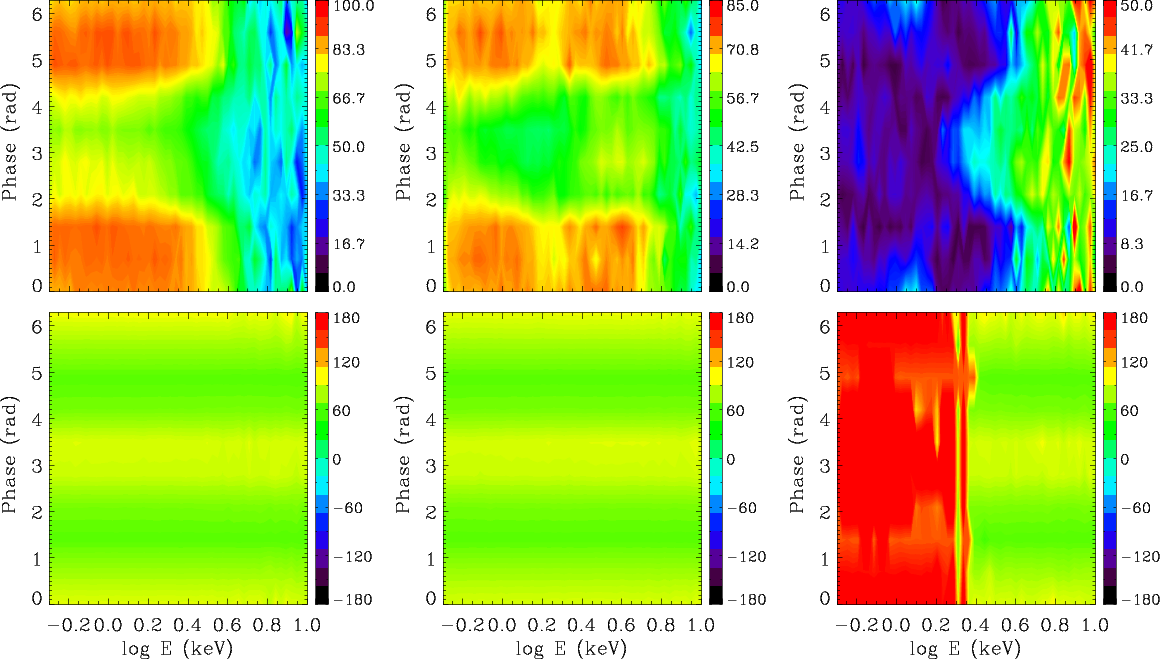}
\caption{Linear polarization degree (top row) and polarization angle (bottom
row) as functions of photon energy and rotational phase for the blackbody
(100\% polarized seed photons, left), atmosphere (middle) and condensed
surface (free-ion limit, right). Note the different color scales for the
polarization degree. The plots refer to $B_{\rm p}=5\times 10^{14}$ G, $
\Deltaphins=0.5$ rad, $\beta=0.34$, $\chi=90^\circ$ and $\xi=60^\circ$.}
\label{figure:pfpacontour}
\end{center}
\end{figure*}

As in our discussion of RCS spectra, we first introduce
the polarization properties of surface thermal radiation,
calculated following the same approach outlined in
\citet{gonz+16}, since this will help in disentangling
the effects produced by resonant scattering from those
intrinsic to surface emission. Figure \ref{figure:PiLmorechi}
shows the energy dependent polarization
fraction $\Pi_{\rm L}$ measured at infinity for the atmosphere
and condensed surface models. Vacuum birefringence is 
included, while GR effects are ignored for consistency with 
the Montecarlo calculation. Atmospheric emission turns out 
to be strongly polarized ($\approx 80$--$100\%$) over the
entire considered energy band. On
the other hand, $\Pi_{\rm L}$ is substantially reduced
for the condensed surface with respect to both the blackbody
and the atmosphere cases, generally attaining a value
smaller than $20\%$. In particular, local minima (where
$\Pi_{\rm L}\sim0$) occur at different photon energies,
according to the different geometry considered. Looking
at the polarization angle behavior (not shown here, but
see the bottom-right panel of Figure \ref{figure:pfpacontour}
for the general picture), these minima correspond to
energies at which the dominant photon polarization state
switches from one normal mode to the other.

\begin{figure*}
\begin{center}
\includegraphics[width=17.5cm]{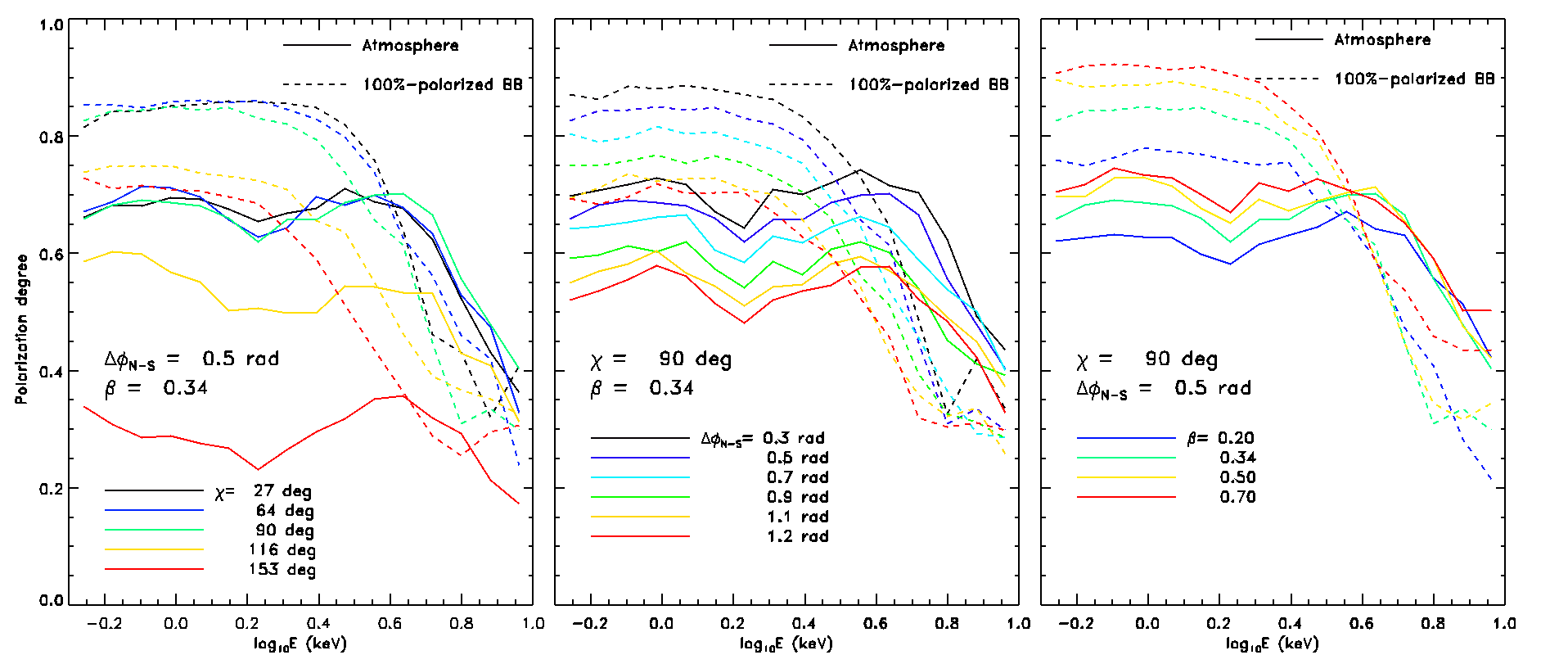}
\caption{Linear polarization degree as a function of the photon energy computed in the case of a magnetized, partially ionized H atmosphere
(solid lines). Details as in Figures \ref{figure:specatmo}--
\ref{figure:specfix}.}
\label{figure:pdatmo}
\end{center}
\end{figure*}

\begin{figure*}
\begin{center}
\includegraphics[width=17.5cm]{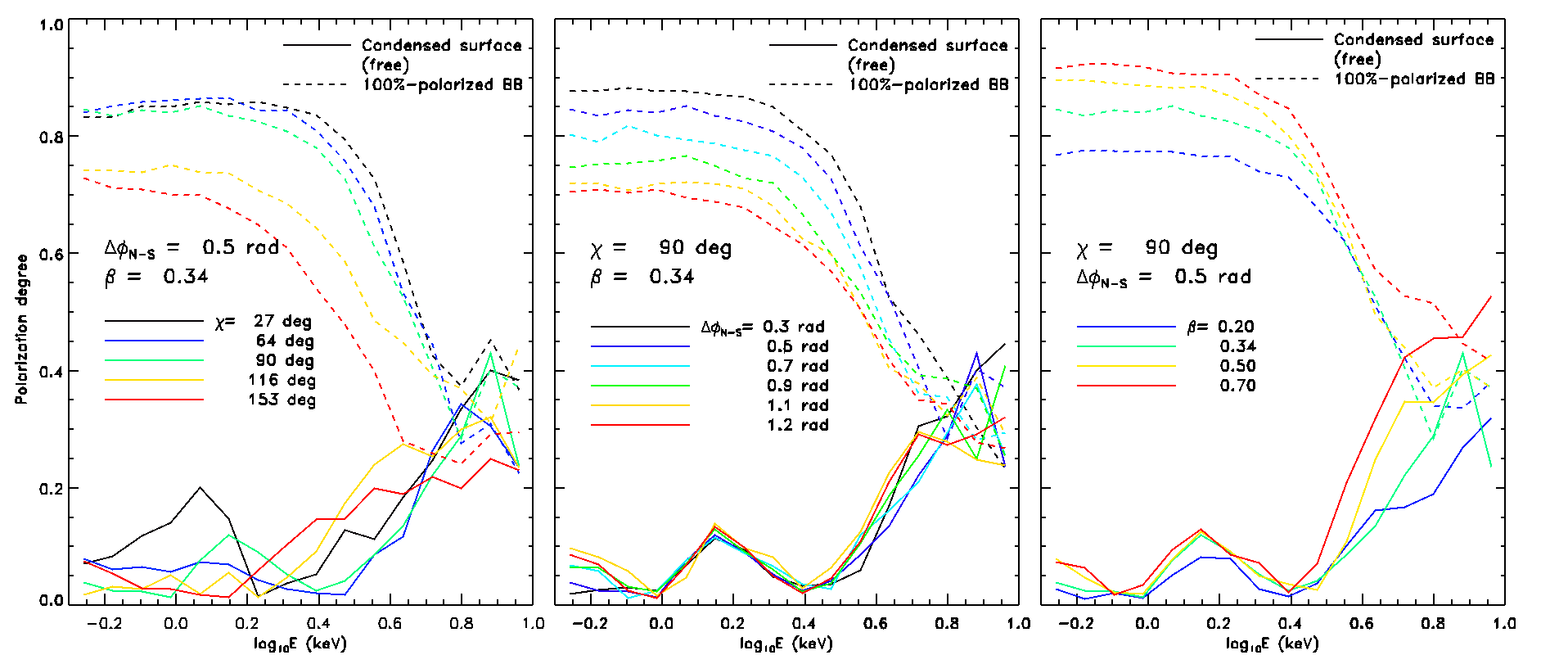}
\caption{Same as in Figure \ref{figure:pdatmo}, but in the case of condensed
surface emission in the free-ion limit.}
\label{figure:pdfree}
\end{center}
\end{figure*}

\begin{figure*}
\begin{center}
\includegraphics[width=17.5cm]{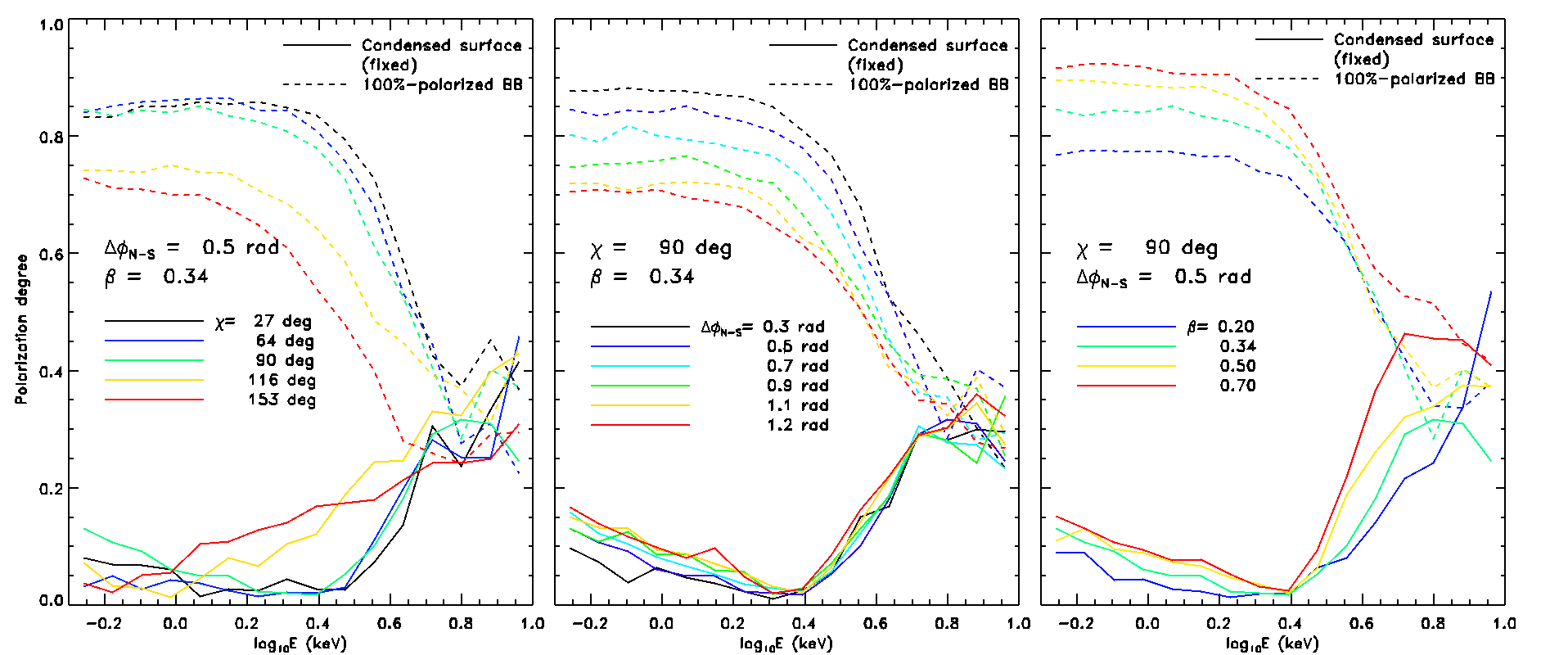}
\caption{Same as in Figure \ref{figure:pdatmo}, but in the case of condensed
surface emission in the fixed-ion limit.}
\label{figure:pdfix}
\end{center}
\end{figure*}

The complete RCS results are reported in figures \ref{figure:pdatmo},
\ref{figure:pdfree} and \ref{figure:pdfix}, following the same scheme discussed
section \ref{subsect:spectra}. Comparing the left-hand panels with the plots
shown in Figure \ref{figure:PiLmorechi}, it can be noted that the behavior of
$\Pi_{\rm L}$ below $\sim 4$ keV bears the imprint of the different intrinsic
polarization patterns. On the other hand, the
polarization degree dramatically changes with respect to the intrinsic one in
the high-energy tail ($\ga 5$ keV), where scattering effects become
dominant. RCS acts in depolarizing radiation in the
atmosphere case, much in the same way as in the case of 100\% blackbody
radiation, since X-mode seed photons are progressively converted into O-mode ones
by scattering \cite[with $33\%$ chance, see][]{ntz08}. The opposite occurs,
instead, for photons emitted by the condensed surface, which acquire more and
more polarization as they scatter in the star magnetosphere\footnote{The rise
in $\Pi_{\rm L}$ at higher energies may be also due to a marginal increase in
the intrinsic polarization fraction, as it can be seen in the middle and right
panels of Figure \ref{figure:PiLmorechi}.}. In particular, for all the
different emission models considered, the polarization degree reaches a value $
\approx20$--$40\%$, in agreement with the expected saturation value, $33\%$,
imposed by the ratio between the X- and O-mode RCS scattering cross
sections.

The polarization degree as a function of the photon energy in the 
magnetized atmosphere case (see Figure \ref{figure:pdatmo}) is fairly
constant al low energies ($\la 4$ keV), exhibiting only a modest decrease 
in correspondence to the proton absorption feature of the primary spectrum.
It then declines at higher energies, although it is somehow higher than 
in the blackbody case for the different configurations. This trend, which 
may appear quite surprising since blackbody photons are $100\%$ polarized 
in the X-mode, can be understood noticing that atmospheric spectra are harder 
than a blackbody (see \S \ref{subsect:spectra}) and that the intrinsic polarization 
still approaches $100\%$ at high energies (see Figure \ref{figure:PiLmorechi}, 
left panel). As mentioned discussing the spectra, while in
the case of blackbody emission the power-law tail is predominantly populated by
low-energy, up-scattered seed photons, there is a still important fraction
of highly-polarized seed photons around $\approx 3$--$4$ keV in the
atmosphere case, which mitigates the effects of RCS maintaining the overall
polarization degree relatively large, before it decreases above $\approx 5$
keV.

The trends of $\Pi_{\rm L}$ as a function of the model parameters 
exhibited in Figures \ref{figure:pdatmo}--\ref{figure:pdfix} can 
be understood as follows. Concerning the LOS inclination (left panels), 
we note that in the blackbody case $\Pi_{\rm L}$ generally decreases
by increasing $\chi$. This is due to the fact that particles are 
assumed to stream from the north to the south magnetic pole (unidirectional 
flow), so that in the southern hemisphere scatterings are mostly head-on, 
resulting in a larger cross section. The same effect is responsible 
for the hardening of the spectrum at high energies seen in the left 
panels of Figures \ref{figure:specatmo}, \ref{figure:specfree} and 
\ref{figure:specfix} \cite[see also][]{ntz08}. This also shows that 
the $0.2$--$2$ keV range contains photons that indeed scatter but 
without a substantial change in their energy (and in fact the spectral 
shape is unchanged, see Figures \ref{figure:specatmo}), together with 
unscattered, primary ones. In this respect, we note that the way 
RCS polarizes radiation is different from that of non-magnetic 
Thomson scattering in which geometry plays a fundamental 
role. The decrease of $\Pi_{\rm L}$ with $\chi$ is less prominent 
in the condensed surface case, since at low 
energies the intrinsic polarization is small, becoming sizeable 
only above $\sim 5$ keV, where up-scattered photons dominate. Actually, 
in the $2$--$4$ keV energy range the trend may be even reversed, 
with the polarization degree consistently higher for $\chi>90^\circ$
than for the other directions. At these intermediate energies, in fact, 
the polarizing effect of RCS starts to be evident as soon as the southern
hemisphere is in view (see also the discussion of Figure \ref{figure:pfpacontour} 
below). In a similar way, the change of the polarization degree with 
$\Deltaphins$ and $\beta$ (see middle and right panels, respectively) 
is related to the form of the RCS optical depth, which roughly goes as 
$\sim\Deltaphins/\beta$ \cite[][]{ft07,ntz08}. In fact, in the blackbody 
case $\Pi_{\rm L}$ decreases by increasing $\Deltaphins$ and decreasing 
$\beta$. In the condensed surface case, instead, this is mirrored only 
at higher energies, where the polarization degree is large enough to make
the effect visible.

The behavior of the polarization degree and angle ($\chi_{\rm
pol}$) as functions of the photon energy and the rotational phase $\gamma$ is
illustrated in Figure \ref{figure:pfpacontour} for 100\% polarized
blackbody (left column), atmosphere (middle column) and condensed surface
(free-ions, right column) models. Here the magnetic axis and the observer's LOS
are at $\xi=60^\circ$ and $\chi=90^\circ$ with respect to the star rotational
axis; the model parameters are fixed at $\Deltaphins=0.5$ rad and $\beta=0.34$.
The variation of $\Pi_{\rm L}$ with the rotational phase can be explained
taking into account that surface patches at different magnetic colatitudes
$\theta$ enter into view as the star rotates, since $\theta$ varies with $
\gamma$ according to $\cos\theta=\cos\chi\cos\xi+\sin\chi\sin\xi\cos\gamma$
\cite[see e.g.][in particular $30^\circ\leq\theta\leq150^\circ$ for the case at
hand]{tav+14}. For 100\% polarized blackbody radiation (top-left
panel), the polarization degree at low photon energies turns out to be as high
as $\sim 80\%$ for $-\pi/2\la\gamma\la\pi/2$ (when $30^\circ\la\theta\la90^
\circ$), while it drops to $\sim 60\%$ at $\gamma\sim\pi$, which corresponds to
$\theta\sim150^\circ$. This is tantamount to say that the northern (magnetic)
hemisphere is in view in the former configuration, whereas the southern one is
visible in the latter. As discussed above (see the left panels of Figures
\ref{figure:pdatmo}--\ref{figure:pdfix}), a higher polarization degree is
expected when the northern magnetic pole enters into view. On the contrary, at high energies 
($\ga 3$ keV), where scattering effects are more significant,
(see the left panels in Figures \ref{figure:specatmo}--\ref{figure:specfix}),
the polarization fraction is below $\approx50\%$ and quite independent of the
rotational phase (despite the low statistics introduces some noise at these energies).

The behaviour of the polarization fraction and angle in the case of a
magnetized atmosphere (middle column) follows closely that of the
blackbody although, as discussed earlier on, the polarization fraction is
somewhat higher at higher energies and lower around the absorption feature. For
the condensed surface model (right column) the polarization fraction at
low photon energies is quite small ($\la10\%$), with in general a fairly weak
dependence on the rotational phase. The only exception occurs in the $2$--$4$
keV energy range, where $\Pi_{\rm L}$ increases for $0<\gamma<\pi$,
attains a maximum ($\approx30\%$) at $\gamma\approx\pi$ and then decreases again
for $\pi\la\gamma\la2\pi$. This peculiar trend can be again explained as an
effect of RCS that, as mentioned above, tends in general to polarize
radiation in the condensed surface case. In fact, the increase in $\Pi_{\rm L}$
due to scatterings becomes sizeable precisely around these energies, starting
at those phases ($\gamma\approx\pi$) at which the southern magnetic hemisphere,
where RCS is more efficient, enters into view (see also the discussion of
Figures \ref{figure:pdfree} and \ref{figure:pdfix} above). At higher energies
the polarization degree settles down to $\approx30$--$40\%$ without an evident
variation with the rotational phase.

The polarization properties for the condensed surface
model just discussed are further confirmed by the behavior
of polarization angle (bottom-right panel). The latter
is characterized by two distinct regimes: at low photon
energies ($\la2$ keV), $\chi_{\rm pol}$ is nearly constant
at $\sim180^\circ$ over the entire phase interval, which
is compatible with a larger fraction of O-mode photons;
at higher energies ($\ga2$ keV), it oscillates around
an average value $\sim90^\circ$, due to a predominance
of X-mode photons \cite[see e.g.][]{tav+15}. The fact
that X-mode photons become dominant starting at $2$--$4$
keV, precisely where the increase of $\Pi_{\rm L}$ occurs,
is again due to the effect of RCS, since scatterings
drive an increase of X-mode photons
with respect to O-mode ones \cite[as already noted by][]{ntz08}.
On the other hand, the polarization angle for both the
blackbody (bottom-left panel) and the atmosphere (bottom-middle
panel) models resembles in the entire energy range the
trend exhibited at high energies for the condensed surface
case, with an overall predominance of extraordinary
photons. Whilst in the blackbody case this is produced
by the choice of 100\% polarized seed photons in the
X-mode, in the atmosphere one it reflects the intrinsic
properties of radiative processes in the presence of
strong magnetic fields, with X-mode opacities much
reduced with respect to O-mode ones.

\section{Simulating observations with {\em IXPE}}
\label{section:ixpe}
Despite polarization measurements are unanimously recognized
as a key tool in modern astrophysics and are routinely
used in different bands of the electromagnetic spectrum,
no dedicated space mission for X-ray polarimetry ever
flew. Since the pioneering time of {\em OSO-8} \cite[][]{weiss+78},
polarimetric measures at X-rays relied only on rocket
and baloon experiments \cite[see][for an overview]{bell+10,weiss18},
with the exception of the small GPD detector {\em PolarLight}
\cite[][]{feng+19}, which flew in 2018
on board of the CubeSat satellite.
Only recently, with the development of photoelectric
polarimeters \cite[][]{costa+01}, the interest in X-ray
polarimetric observatories was revived. This led to a
number of proposals, including the NASA {\em GEMS}
\cite[][]{baum+12} and {\em PRAXyS} \cite[][]{jah+14},
and the ESA {\em XIPE} \cite[][]{soff+16}, which however
were not selected. Nowadays, two promising missions are
looming on the horizon: the {\em IXPE} mission \cite[][]
{weiss+16}, selected for the NASA SMEX programme and 
scheduled for launch in early 2021; and the {\em eXTP} 
mission \cite[][]{zhang+19,sant+19b}, which has been approved by 
the Chinese Academy of Science, and is expected to fly in 
2025. 

Magnetars are among the key science targets of {\em IXPE}
and {\em eXTP}, especially in connection with the possibility
to test birefringence in the ultra-magnetized vacuum around
NSs \cite[][]{sant+19a}. On the wake of the results discussed 
in \citet{tav+14}, here we present a set of simulated {\em 
IXPE} observations meant to be representative of the 
bright AXPs 1RXS J170849.0$-$400910 and 4U 0142$+$61, which 
have been selected for the first year of operations. The model 
parameters are $\Deltaphins=0.5$ rad, $\beta=0.34$ and $T=0.5$ 
keV (constant over the surface), i.e. the same derived from 
the RCS spectral fitting of J1708 \cite[][]{zane+09}, and $B_
{\rm p}=5\times10^{14}$ G. The different flux level of the two 
sources is accounted for by rescaling the Montecarlo output in such
a way to produce a $2$--$10$ keV unabsorbed flux of $2.4$ ($6.8$)
$\times10^{-11}\,\flux$ \cite[see again][]{ok14}.

Phase-dependent simulations for the flux, linear polarization
fraction and angle were produced using a specific code \cite[see
e.g.][for more information]{tav+14} which incorporates {\em IXPE}
instrumental setup (modulation factor and effective area according
to the most updated release). The phase bins were selected
in such a way to ensure that each bin contains a large enough
number of counts. In the case of J1708 (4U 0142$+$61) we checked 
that, with 9 bins, about $5$ ($13$) photons are collected in any
phase interval. For each simulated observation, we performed 
$N\sim100$ realizations based on the
same input model ($\chi=90^\circ$, $\xi=60^\circ$, $\Deltaphins=0.5$, $
\beta=0.34$). The flux, polarization degree and angle derived from
each of these were simultaneously fitted using
an archive of models obtained varying the parameters; the one
with the median reduced $\chi^2$ is finally selected.

\begin{figure*}
\begin{center}
\includegraphics[width=17.5cm]{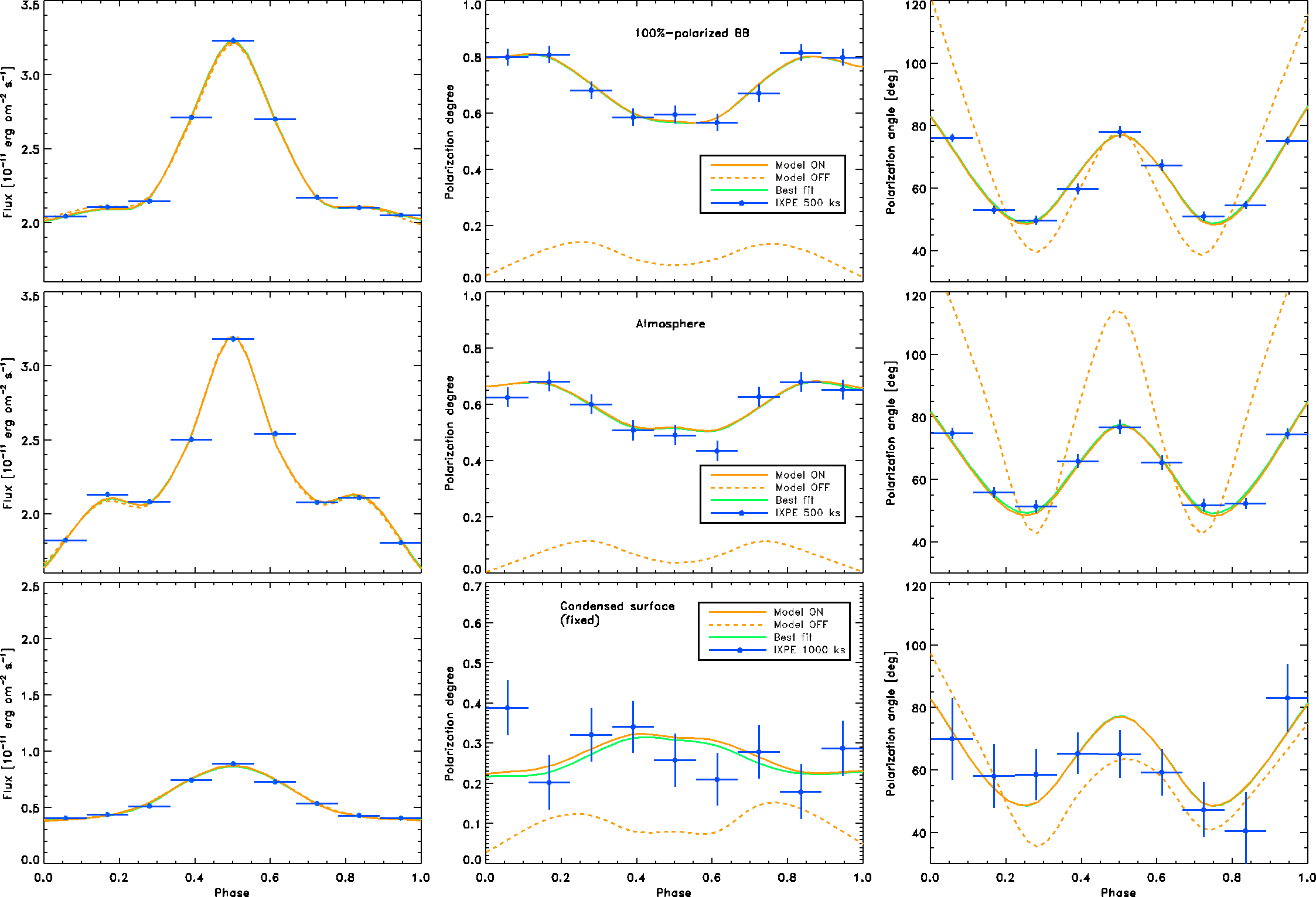}
\caption{Flux (left column), linear polarization degree (middle
column) and polarization angle (right column) as functions of
the rotational phase for blackbody (100\% polarized seed
photons, top row), atmosphere (middle row) and condensed surface
(fixed-ion limit, bottom row). The model parameters are 
$B_{\rm p}=5\times10^{14}$ G, $\Deltaphins=0.5$ rad, $\beta=0.34$,
$\chi=90^\circ$ and $\xi=60^\circ$.
The surface temperature and the $2$--$10$ keV unabsorbed flux 
are $0.5$ keV and $2.4\times10^{-11}$ erg cm$^{-2}$ s$^{-1}$,
respectively, as appropriate for J1708. Filled circles with error bars
(at $1\sigma$) are the simulated data. The orange and green solid
lines show the model from which data were obtained and the best
simultaneous fit, respectively. The orange dashed lines refer
to the QED-OFF case (see text for more details).}
\label{figure:fittotal}
\end{center}
\end{figure*}
\begin{figure*}
\begin{center}
\includegraphics[width=17.5cm]{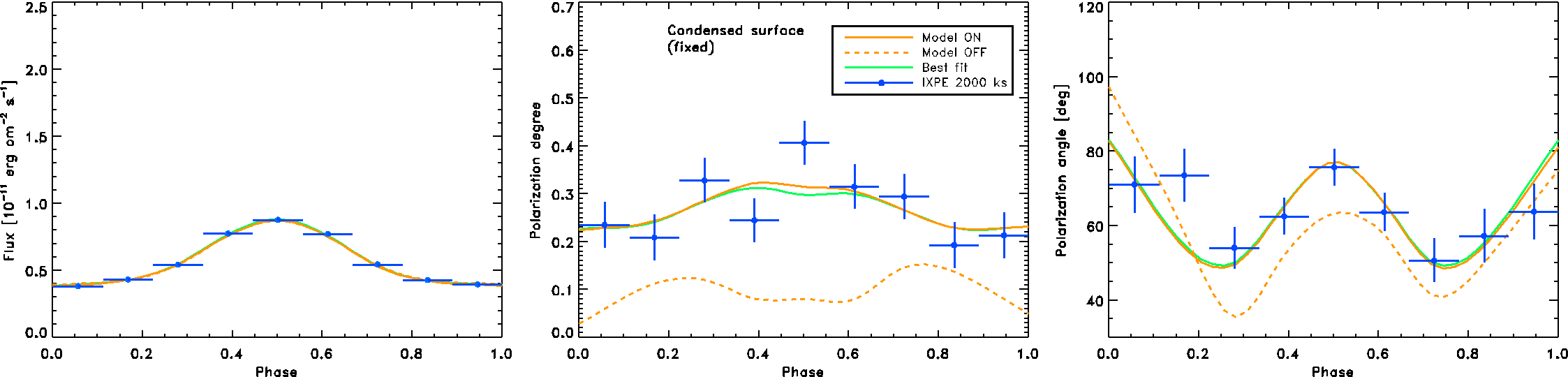}
\caption{Same as in Figure \ref{figure:fittotal} (bottom row),
but for $t_{\rm exp}=2$ Ms.}
\label{figure:fitfix2000}
\end{center}
\end{figure*}

\begin{table*}
\begin{center}
\begin{tabular}{lllllll}
\hline
\  & $t_{\rm exp}$ (ks) & $\ \ \ \ \ \ \chi$ (deg) & $\ \ \ \ \ \xi$ (deg) & $\ \Deltaphins$ (rad) & $\ \ \ \ \ \ \ \ \ \ \ \beta$ & $\chi^2_{\rm red}$ \\
\hline
Input values &\ \ \ -- & \ \ \ \ \ \ \ \ \ \ \ 90 & \ \ \ \ \ \ \ \ \ 60 & \ \ \ \ \ \ \ \ \ 0.5 &\ \ \ \ \ \ \ \ \ 0.34 &\ \ \ -- \\
\hline
100\% polarized BB & \ &\ & \ & \ & \ & \ \\
QED-ON & $\ 500$ & \ \ $90.80\pm1.27$ & $59.53\pm0.90$ & $0.502\pm0.019$ & $0.340\pm0.008$ & $1.14$ \\
QED-OFF & $\ 500$ & $102.51\pm1.17$ & $49.80\pm0.79$ & $0.300\pm0.000^{\mathrm a}$ & $0.411\pm0.005$ & $137.1$ \\
\hline
Atmosphere & \ &\ & \ & \ & \ & \ \\
QED-ON & $\ 500$ & \ \ $90.35\pm1.01$ & $59.28\pm1.01$ & $0.507\pm0.025$ & $0.334\pm0.013$ & $1.58$ \\
QED-OFF & $\ 500$ & \ \ $91.31\pm0.71$ & $56.30\pm0.73$ & $0.348\pm0.011$ & $0.445\pm0.007$ & $103.2$ \\
\hline
Condensed surface (fixed-ions) & \ &\ & \ & \ & \ & \ \\
QED-ON & $1000$ & \ \  $88.81\pm7.27$ & $60.85\pm4.31$ & $0.522\pm0.053$ & $0.324\pm0.026$ & $1.13$ \\
QED-OFF & $1000$ & \ \ $93.74\pm6.71$ & $65.96\pm2.70$ & $0.450\pm0.021$ & $0.355\pm0.011$ & $2.38$ \\
\hline
Condensed surface (fixed-ions) & \ &\ & \ & \ & \ & \ \\
QED-ON & $2000$ & \ \ $91.89\pm4.64$ & $59.57\pm2.44$ & $0.511\pm0.030$ & $0.337\pm0.017$ & $1.16$ \\
QED-OFF & $2000$ & \ \ $98.20\pm3.44$ & $53.65\pm1.79$ & $0.468\pm0.017$ & $0.354\pm0.011$ & $3.95$ \\
\hline
\end{tabular}
\caption{Results of the simultaneous fits of the flux, linear polarization degree 
and polarization angle shown in Figures \ref{figure:fittotal} and \ref{figure:fitfix2000}, 
in the case of 100\% polarized blackbody, atmosphere and condensed surface (fixed-ions) 
models, either with QED effects (QED-ON) or without (QED-OFF). The reduced $\chi^2$ for
the OFF models is the minimum among the different realizations. Reported errors are at 
the $1\sigma$ level. \newline
$^{\mathrm a}$ During the fit this parameter hit the lower bound of the range.}
\label{table:fits}
\end{center}
\end{table*}

Figure \ref{figure:fittotal} illustrates the results for 100\% polarized blackbody (top row), atmosphere
(middle row) and condensed surface (fixed-ions\footnote{Results for free-ions are very similar and are not shown.},
bottom row) emission models. Simulated data refer to the $2$--$8$ keV energy range for both the blackbody and
atmosphere, and to the $4$--$8$ keV range for the condensed surface, for which the contribution to $\Pi_{\rm L}$
between $2$ and $4$ keV is small (see Figure \ref{figure:pdfix}). An exposure time $t_{\rm exp}=500$ ks is assumed,
apart from the condensed surface model, for which $t_{\rm exp}=1$ Ms. The orange solid line represents the model from
which the mock data were generated, while the green solid line corresponds to the best fitting model. In all cases
the fit recovers the input parameters with good accuracy (within $1\sigma$, see Table \ref{table:fits}). The dashed
orange line shows instead the input model computed without accounting for the vacuum polarization effects (QED-OFF).
With the selected $t_{\rm exp}$, the QED-OFF model is always rejected with high confidence, with typical values of
reduced $\chi^2\ga 4$ vs $1.13$--$1.58$. We note that the present treatment of the QED-OFF
case differs slightly from that described in \citet{tav+14} inasmuch the adiabatic radius is artificially moved to
the last-scattering 
radius/star surface for photons which do/do not undergo scatterings.
To all effects this results
in an overestimate of the polarization degree calculated in the QED-OFF case. With reference to Figure
\ref{figure:fittotal}, a polarization measure $\ga 20\%$ will be sufficient to validate vacuum birefringence effects in
strong magnetic fields.

\begin{figure*}
\begin{center}
\includegraphics[width=17.5cm]{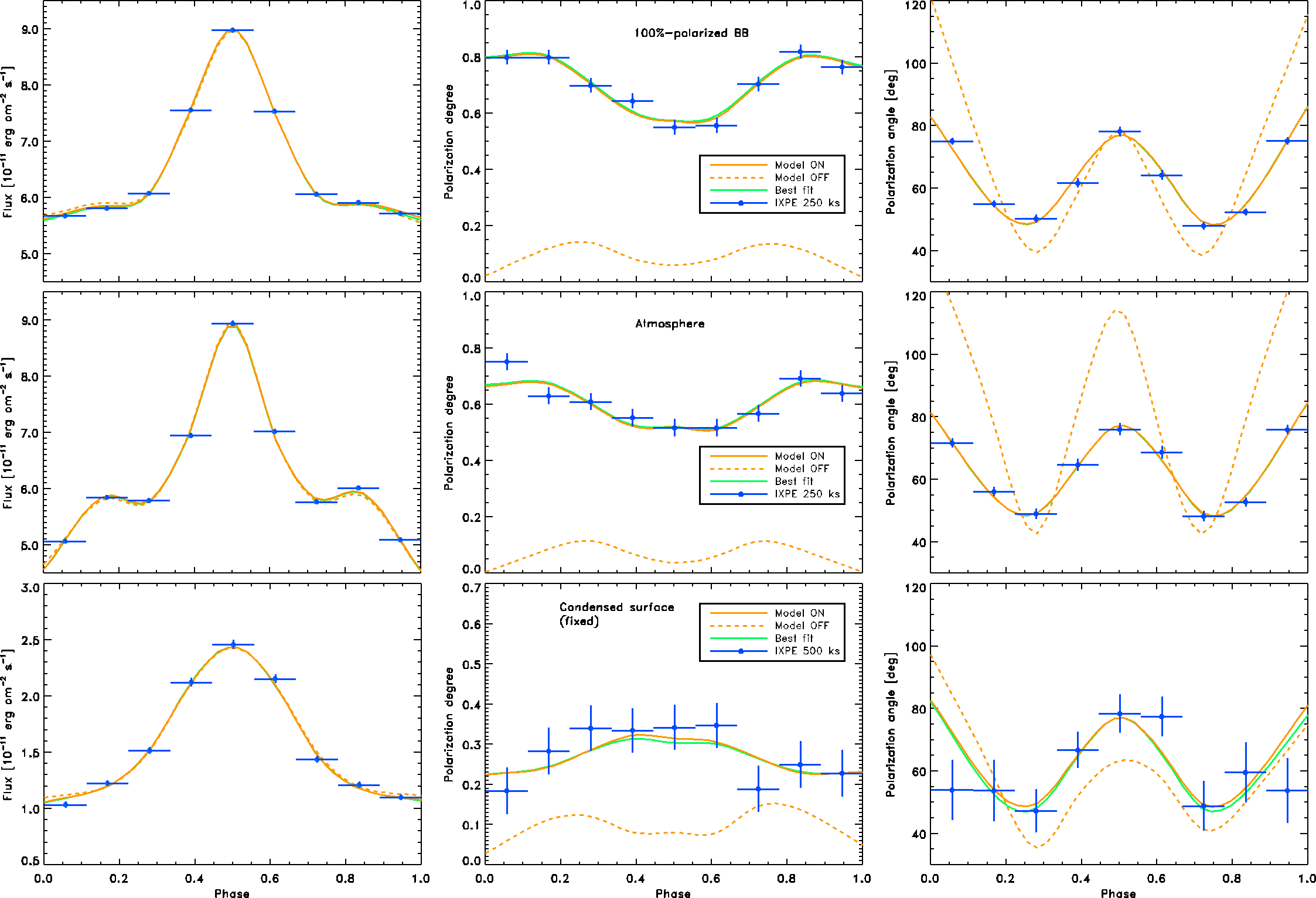}
\caption{Same as in Figure \ref{figure:fittotal} but with $2$--$10$
keV unabsorbed flux of $6.9\times10^{-11}$ erg cm$^{-2}$ s$^{-1}$
as for AXP 4U 0142$+$61. An
exposure time of $250$ ks has been considered for blackbody and
atmosphere mock data, while $t_{\rm exp}=500$ ks is taken for
the condensed surface case.}
\label{figure:fittotal4U}
\end{center}
\end{figure*}

Because of the comparatively low polarization degree expected in the condensed
surface models, the errors on both the polarization fraction and angle are much 
larger than in the blackbody and the atmosphere cases. In order to verify if 
results are strongly dependent on the exposure time, we produced a further 
(fixed-ion) simulation with $t_{\rm exp}=2$ Ms (see Figure 
\ref{figure:fitfix2000}). While the error bars are sensibly reduced and the 
robustness of the QED-ON vs QED-OFF fits improves, the model parameters turn 
out to be fully compatible with those of the 1 Ms simulation (see Table 
\ref{table:fits}). Figure \ref{figure:fittotal4U} shows the same simulations as 
in Figure \ref{figure:fittotal}, but assuming the $2$--$10$ keV unabsorbed 
flux of the AXP 4U 0142$+$61 (i.e. $\sim3$ times higher than that used 
previously). As expected, results in this case (see Table \ref{table:fits4U})
are fully compatible with those obtained previously; in 
particular, the higher value of the flux allows us to reduce the exposure time 
(by a factor of $\sim2$) and still get the same level of accuracy.
\begin{table*}
\begin{center}
\begin{tabular}{lllllll}
\hline
\  & $t_{\rm exp}$ (ks) & $\ \ \ \ \ \ \chi$ (deg) & $\ \ \ \ \ \xi$ (deg) & $\ \Deltaphins$ (rad) & $\ \ \ \ \ \ \ \ \ \ \ \beta$ & $\chi^2_{\rm red}$ \\
\hline
Input values &\ \ \ -- & \ \ \ \ \ \ \ \ \ \ \ 90 &  \ \ \ \ \ \ \ \ \ 60 & \ \ \ \ \ \ \ \ \ 0.5 &\ \ \ \ \ \ \ \ \ 0.34 &\ \ \ -- \\
\hline
100\% polarized BB & \ &\ & \ & \ & \ & \ \\
QED-ON & $\ 250$ & \ \ $90.61\pm1.13$ & $59.73\pm0.78$ & $0.489\pm0.016$ & $0.349\pm0.007$ & $1.16$ \\
QED-OFF & $\ 250$ & $103.68\pm0.97$ & $48.48\pm0.65$ & $0.300\pm0.000^{\mathrm a}$ & $0.412\pm0.004$ & $191.3$ \\
\hline
Atmosphere & \ &\ & \ & \ & \ & \ \\
QED-ON & $\ 250$ & \ \ $90.01\pm0.87$ & $59.92\pm0.86$ & $0.491\pm0.022$ & $0.347\pm0.011$ & $1.53$ \\
QED-OFF & $\ 250$ & \ \ $91.82\pm0.59$ & $55.38\pm0.62$ & $0.350\pm0.009$ & $0.450\pm0.006$ & $140.2$ \\
\hline
Condensed surface (fixed-ions) & \ &\ & \ & \ & \ & \ \\
QED-ON & $\ 500$ & \ \ $87.85\pm6.74$ & $62.06\pm3.53$ & $0.502\pm0.031$ & $0.341\pm0.017$ & $1.14$ \\
QED-OFF & $\ 500$ & \ \ $99.85\pm4.33$ & $54.49\pm1.95$ & $0.433\pm0.025$ & $0.376\pm0.015$ & $3.01$ \\
\hline
\end{tabular}
\caption{Same as in Table \ref{table:fits} but for the case shown in Figure \ref{figure:fittotal4U}.}
\label{table:fits4U}
\end{center}
\end{table*}

\begin{figure*}
\begin{center}
\includegraphics[width=11.7cm]{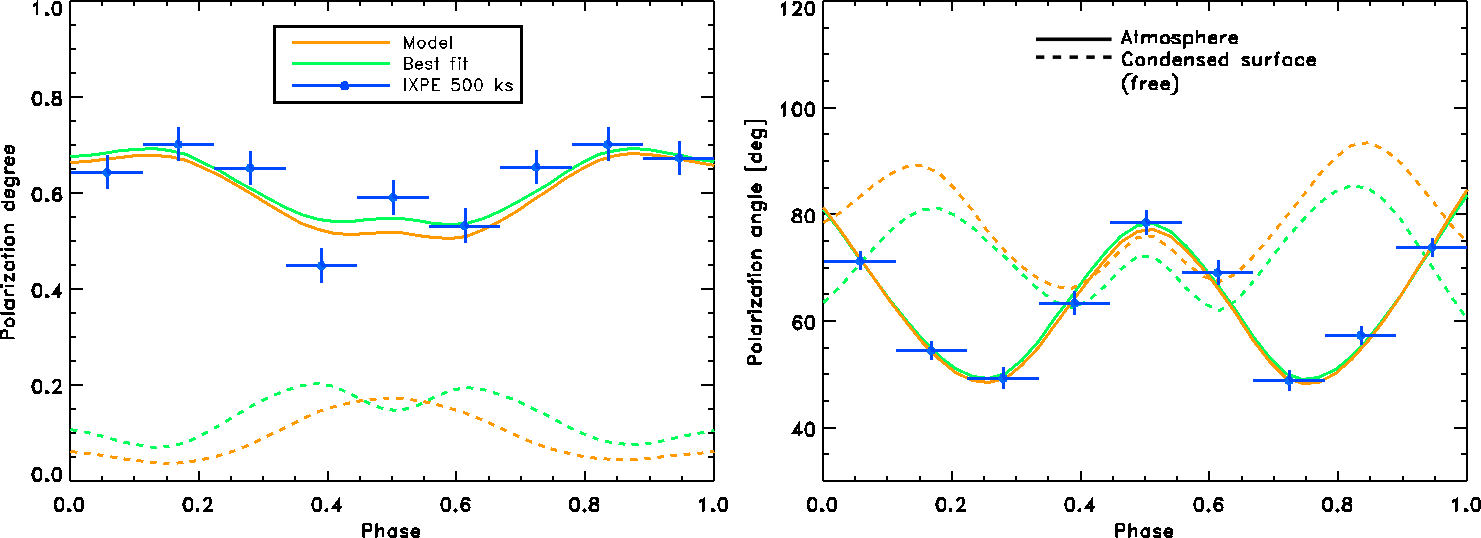}
\caption{Linear polarization degree (left) and angle (right) as functions of 
the rotational phase for $\Bp=5\times10^{14}$ G, $\Deltaphins=0.5$ rad,
$\beta=0.34$, $\chi=90^\circ$, $\xi=60^\circ$ and the $2$--$10$ keV unabsorbed 
flux of J1708 ($t_{\exp}=500$ ks). Filled 
circles with error bars (at $1\sigma$) are the simulated data obtained from 
the atmosphere model in the $2$--$8$ keV band (orange solid line); the 
condensed surface (free-ion) model for the same values of the parameters is 
shown by the orange dashed curve. The green lines represent the best 
simultaneous fits for the atmosphere (solid) and the condensed surface (dashed) 
models.}
\label{figure:atmofreecomp}
\end{center}
\end{figure*}

Given the vastly different expectations for the polarization degree and angles from the different
models discussed above, phase-resolved polarimetry may provide a direct way to probe the physical 
state of a magnetar surface. To test this, we produced synthetic data from the atmosphere model and
fit them both with the atmosphere archive and the condensed surface one (free-ions). Results are
reported in Figure \ref{figure:atmofreecomp}. While the input model parameters are well recovered
by the fit ($\chi^2_{\rm red}=1.17$) in the former case, the agreement between the data and the best
condensed surface model is largely unacceptable ($\chi^2_{\rm red}=160.6$, see Table \ref{table:atmofreecomp}). 
This result is robust and shows that there is no degeneracy in the model fit at variance with what 
occurs considering phase averaged polarization observables \cite[see][]{tav+15,gonz+16}. 

\begin{table*}
\begin{center}
\begin{tabular}{llllll}
\hline
\  & $\ \ \ \ \ \ \chi$ (deg) & $\ \ \ \ \ \xi$ (deg) & $\ \Deltaphins$ (rad) & $\ \ \ \ \ \ \ \ \ \ \ \beta$ & $\chi^2_{\rm red}$ \\
\hline
Input values & \ \ \ \ \ \ \ \ \ 90 &  \ \ \ \ \ \ \ \ \ 60 & \ \ \ \ \ \ \ \ \ 0.5 &\ \ \ \ \ \ \ \ \ 0.34 &\ \ \ -- \\
\hline
{\sc fitting models} &\ & \ & \ & \ & \ \\
Atmosphere & $87.54\pm2.92$ & $58.25\pm1.93$ & $0.470\pm0.061$ & $0.342\pm0.081$ & $1.19$ \\
Condensed surface (free-ions) & $92.34\pm1.38$ & $75.44\pm1.41$ & $0.920\pm0.052$ & $0.200\pm0.000^\mathrm{a}$ & $165.1$ \\
\hline
\end{tabular}
\caption{Results of the simultaneous fits of the linear polarization degree and polarization
angle shown in Figure \ref{figure:atmofreecomp}. The reduced $\chi^2$ for the condensed surface 
model is the minimum among the different realizations. Reported errors are at the $1\sigma$ level.}
\label{table:atmofreecomp}
\end{center}
\end{table*}

\section{Discussion and conclusions}
\label{section:discussion}
In this work the spectral and polarimetric properties of persistent
X-ray emission from magnetar sources have been investigated within the resonant
Compton scattering (RCS) paradigm, first suggested by \citet{tlk02}.
The present study improves over previous ones \cite[][]{ft07,ntz08,fd11,
tav+14} inasmuch more physical surface emission models have been considered:
a magnetized hydrogen atmosphere and a condensed surface. The feasibility of polarization measurements in magnetars 
with the forthcoming X-ray polarimeter {\em IXPE} was also 
readdressed. Our main findings are as follows:
\begin{itemize}
\item while the spectral properties are not very sensitive to the
surface emission model (showing always a ``blackbody$+$power-law'' shape), the 
polarization pattern is strongly affected by the primary photon
spectrum;
\item atmospheric models exhibit a high polarization degree in the X-mode
(up to $\sim 80\%$), depending on the geometrical and magnetospheric 
parameters, not much below that of $100\%$ polarized blackbody seed
photons;
\item condensed surface models (both in the free- and fixed-ion limits)
are much less polarized (up to $\sim 30\%$) and O-mode photons dominate
in the low-energy range;
\item RCS acts in depolarizing the (highly polarized) atmospheric
seed photons as they propagate in the magnetosphere, while the opposite
occurs for the (weakly polarized) condensed surface ones; as a
consequence, at large energies, the polarization degree approches 
in any case $33\%$ (the value expected from the RCS cross sections)
but it does so either from above or from below, respectively;
\item simulations of {\em IXPE} response show that an exposure time
of $0.5$--$1$ Ms is sufficient to measure the polarization observables
in a bright magnetar source, with high enough accuracy to probe (i)
vacuum birefringence and (ii) the physical state of the NS surface.
\end{itemize}

Previous results were obtained under a number
of assumptions, concerning the description of both the star magnetosphere
and the surface temperature distribution. In addition, no general
relativistic effects \cite[see e.g.][]{fd11,tav+15} have been 
accounted for. 

Crustal displacements in a magnetar most likely produce a localized
twist, confined to a bundle of current-carring field lines. Moreover, 
the twist itself decays, shrinking towards the polar regions. 
As a consequence, the star surface should exhibit hot spot(s) 
in correspondence to the footpoints of the twisted field lines, where 
returning currents dissipate, while the rest of the surface is at a 
lower temperature. The size of the hot spot(s) decreases as the magnetosphere 
untwists \cite[][]{belo09}. Actually, our Montecarlo code can handle 
quite general magnetic field configurations, so that in principle it 
would have been possible to treat a localized twist \cite[see e.g.][]{vig+12}.
Also the surface temperature can be specified without restrictions, 
assigning a different value to each patch \cite[see][]{ntz08}. 

In this work, however, we used a simpler description in which the magnetosphere 
is globally twisted (i.e. the entire north magnetic hemisphere is twisted 
wrt the south one), the current flow is unidirectional (i.e. charge carriers
are electrons) and the temperature is constant over the entire surface.
The reason for this choice is twofold: first, it allows for a direct comparison
with previous works \cite[][]{ft07,ntz08,fd11,tav+14}, as already mentioned
in \S \ref{subsection:thermalemission}, and second, it avoids to introduce
a large number of free parameters, which would have been necessary to 
describe a more complex magnetospheric structure and temperature profile. 

The charge carriers flowing along the closed field lines are expected
to be mostly electrons and positrons \cite[][]{bt07}, with a (spatially) 
changing Lorentz factor, as discussed in some more detail by \citet{belo13}.
In the lack of a complete model, which is still to come, we retained both 
the unidirectional flow hypothesis and the assumptions that charges move
with a constant speed.

The surface temperature distribution in a magnetar
is governed by a number of different and competing effects, which are
still not completely understood. Under magnetar conditions electron 
transport in the envelope, which can be assumed to be the main
responsible for heat conduction, occurs almost entirely
along the magnetic field lines. This produces a substantial
temperature variation in going from the magnetic poles to the equator
(see e.g. Potekhin, Pons \& Page \citeyear{pot+15}),
which can potentially modify both the spectral and 
polarization properties of magnetar persistent radiation. 
On the other hand, heating by returning currents can dominate 
over the internal one, making the equatorial belt hotter \cite[][]{bt07}. 
In the light of this, an isothermal surface distribution may be
not too far from the realistic one, and will in any case 
catch the essential features of the model. We stress again
that the effects of returning currents on the surface star
layers are ignored. 

In the case of thermally-emitting NSs, vacuum birefringence strongly
influences the polarization properties at infinity. QED effects are actually 
quite sensitive to the extent of the radiating region on 
the star surface or, equivalently, on the steepness of the 
temperature gradient. As discussed in \citet{vanadpern09}, if emission in these 
sources comes from a small (point-like) polar cap, the depolarization 
due to geometrical effects is not present, so that expectations 
from QED-ON and QED-OFF models are quite the same. In order to 
test vacuum birefringence through the measured polarization degree,
the cap aperture should be $\ga 40^\circ$ \cite[][]{gonz+17,sant+19b}.
Clearly, in magnetars RCS acts in spatially redistributing 
surface photons. If indeed primary photons come from a small
cap, we expect the QED-ON and QED-OFF models to produce nearly 
the same results only (or mostly) below $\sim 2$ keV, where unscattered 
radiation dominates.

Predictions for the polarization observables in the atmospheric 
case are quite dependent on the adopted treatment of the mode
switching at the vacuum resonance, since this strongly influences
the opacities. As a matter of fact, at the mode collapse points 
(vacuum resonances) the normal modes approximation breaks down 
and radiative transfer should be solved using the Stokes 
parameters \cite[see][]{ps79,lh03}. The two limiting cases of 
full mode conversion and no mode conversion have been investigated 
\cite[][]{hl03} and both lead to a large polarization of radiation emerging from 
a single patch ($\sim100\%$ in the soft X-ray band).
In this work we assumed adiabatic mode conversion at the resonance 
and treated the mode switching following \citet{vanadlai06}. Although 
their approach is formally energy and angle dependent, under the 
conditions typical of the atmospheres presented here, most photons 
actually undergo mode switching at the resonance. On the other hand, 
the critical angle at which the normal modes break down is strongly 
dependent on photon frequency and direction \cite[][]{ps79}, 
so that the differences in the X- and O-mode opacities induced 
by the strong field are much reduced when partial mode switching
occurs. This may lead to a much smaller degree of polarization \cite[see
also][for a detailed discussion in the gray case]{gonz+19}.

\section*{Acknowledgments}
We thank Jeremy Heyl for some helpful discussions and an anonymous
referee for his/her constructive criticism which helped in improving
a previous version of this paper.
RT and RT aknowledge financial support from the Italian MIUR through
PRIN grant 2017LJ39LM.
The work of AYP was supported by RFBR and DFG within the research project
19-52-12013.
The work of VS was supported by the DFG grant WE1312/51-1 and the  
Russian Science Foundation grant 19-12-00423.

\label{lastpage}

\end{document}